\documentclass[aps,prd,preprintnumbers,nofootinbib,onecolumn,a4paper]{revtex4-2}%
\usepackage{bm,latexsym,amsmath,amssymb,amsfonts,mathtools,caption}
\usepackage{amssymb, amsmath}
\usepackage[dvipsnames]{xcolor}
\usepackage{hyperref}
\usepackage{booktabs}
\usepackage{mathrsfs}
\usepackage{ulem}
\makeatletter
\newcommand{\figcaption}[1]{\def\@captype{figure}\caption{#1}}
\newcommand{\tblcaption}[1]{\def\@captype{table}\caption{#1}}

\newcommand{\YK}[1]{{\color{red} #1}}

\newcommand{\UK}[1]{{\color{magenta} #1}}
\newcommand{\UKsout}[1]{{\color{magenta} \sout{#1}}}

\makeatother
\begin{document}
\title{Entanglement of the Vacuum between Left, Right, Future, and Past: \\Dirac spinor in Rindler spaces and Kasner spaces}

\author{Kazushige Ueda$^1$}
\author{Atsushi Higuchi$^2$}
\author{Kazuhiro Yamamoto$^{1,3}$}
\author{Ar Rohim$^4$}
\author{Yue Nan$^4$}

\affiliation{$^1$Department of Physics, Kyushu University, 744 Motooka, Nishi-Ku, Fukuoka 819-0395, Japan}
\affiliation{$^2$Department of Mathematics, University of York, Heslington, York YO10 5DD, United Kingdom}
\affiliation{$^3$Research Center for Advanced Particle Physics, Kyushu University, 744 Motooka, Nishi-ku, Fukuoka 819-0395, Japan}
\affiliation{$^4$Graduate School of Science, Hiroshima University, 1-3-1 Kagamiyama, Higashi-Hiroshima, Japan  739-8526}

\begin{abstract} 
We study the relations of the positive frequency mode functions of Dirac field in 4-dimensional Minkowski spacetime covered with Rindler and Kasner coordinates, and describe the explicit form of the Minkowski vacuum state with the quantum states in Kasner and Rindler regions, and analytically continue the solutions. As a result, we obtain the correspondence of the positive frequency mode functions in Kasner region and Rindler region in a
unified manner which derives vacuum entanglement.
\end{abstract} 
\maketitle

\section{Introduction}
\label{Introduction}
Quantum entanglement is one of the most important features of quantum physics, which is a concept essential to quantum information theory, technology, and related topics.  Quantum entanglement plays an important role not only in discussions of quantum optics but also in discussions of quantum field theory in curved spacetime, as exemplified by the Unruh effect and the Hawking effect. The research into these effects of quantum fields in curved spacetime might give us a hint for unifying the gravity theory and quantum mechanics.
The Unruh effect is the well-known theoretical prediction that a uniformly accelerated observer (Rindler observer) sees the vacuum state in an inertial frame (Minkowski vacuum state) as a thermally excited state characterized by the temperature proportional to the acceleration $a$ of the Rindler observer \cite{Unruh,UnruhWald}. Topics related to the Unruh effect have been studied well due to its importance and simplicity (see, e.g., \cite{Higuchi} for a review). 

To demonstrate the Unruh effect, various experiments have been proposed \cite{Bell,Bell2,Akhmedov,Akhmedov2,Barber,UnruhPR,Mumller,Vanzella,Vanzella2,Matsas2}.
One of the big questions in these topics is whether a uniformly accelerated object coupled to a quantum field (the Unruh-de Witt detector) would emit radiation or not \cite{ChenTajima,Schutzhold,Schutzhold2}. This point is carefully discussed in many works (see, e.g., Refs.~\cite{SNRKS,LinHu,HuRaval}). Authors of Ref.~\cite{HuRaval} showed that the radiation would be cancelled in the 1+1 dimensional case.  This result would appear to agree with our intuition that there is no radiation when the detector is in a thermal equilibrium state. However, the analysis of the 1+3 dimensional case \cite{SNRKS,ITUY,HIUY} gives a counter-intuitive result: there is some quantum radiation induced by a non-local correlation of the quantum field in the Minkowski vacuum state due to the vacuum entanglement between the left (L) and right (R) Rindler wedges. 
The Minkowski vacuum state of a scalar field is described by an entangled state between the quantum states on the left (L) and right (R) Rindler wedges~\cite{Unruh, UnruhWald} heuristically expressed as follows:
\begin{eqnarray}
|{\rm 0, M}\rangle \propto \prod_{j} \biggr[ \sum_{n_j=0}^{\infty} e^{-\pi n_j \omega_j/a}| n_j \rangle_{\rm R} \otimes | n_j \rangle_{\rm L} \biggl].
\label{MVacuum}
\end{eqnarray}%
Here $|n_j\rangle_{\rm R(L)}$ is the $n_j$-th excited state on the R(L) Rindler region (see Fig. \ref{Mcoordinate}) with an acceleration $a$, and $j=(\omega,\bm{{\bm k}_\perp})$ schematically denotes a mode specified by energy $\omega$ and momentum ${\bm k}_{\perp}$ perpendicular to the direction of acceleration.  The roman letters ``R" and ``L" denote the right Rindler region and the left Rindler region, respectively.
This expression is often used to describe the Unruh effect. Because a uniformly accelerated observer in the R region is causally disconnected to the events in the L region, we take the partial trace of the density operator with respect to the Hilbert space of the L region to obtain the reduced density operator in the R region. 
This leads to a reduced density operator representing the thermal state at the Unruh 
temperature $T_{\rm U}= a/2\pi$. 

Some of the authors of the present paper examined the description of the Minkowski vacuum state of a scalar field to extend the expression Eq.~(\ref{MVacuum}) to the future (expanding) Kasner spacetime (F region) and the past (shrinking) Kasner spacetime (P region). This is done by analytic continuation of the mode functions in the F(P) region into the R region and the L region~\cite{HIUY}. This work yielded a result generalizing the work by Olson and Ralph~\cite{OlsonRalph}, which claimed a time-like entanglement in the F region and the P region in the case of the two-dimensional massless scalar field to that in the general four-dimensional case. 
Furthermore, the authors of Refs.~\cite{HIUY,ITUY} analyzed the quantum radiation produced by a uniformly accelerated Unruh-de~Witt detector coupled to a scalar field. The property of the quantum radiation is entirely different from the usual radiation locally generated: the quantum radiation is interpreted to be induced by the quantum entanglement, which is behind the Unruh effect.  
The entanglement-based description of the Minkowski vacuum state is important for analyzing the theoretical predictions for the quantum radiation associated with the Unruh effect. It is important to extend the analysis of the Minkowski vacuum entanglement to the whole region,  including the F and P regions since the observer in the F region will receive information from the R region where the uniformly accelerated object is. When we consider experimental models related to quantum radiation, the relation between the modes in the four quadrants of the field needs to be taken into account.

In the present paper, we consider the entanglement-based description of the
Minkowski vacuum state of a Dirac field.  
Some papers in the literature investigate a Dirac field in Rindler spacetime~\cite{Vanzella,Higuchi,McMahon,Alsing.et.al}; Ref.~\cite{McMahon} treats this field in Kasner spacetime. These papers discussed the entanglement-based description of the Minkowski vacuum state 
of the Dirac field with the use of the quantum states in Rindler spacetime. 
Alsing {\it et al.} found that the entanglement between two Rindler modes of a free Dirac field is degraded by the Unruh effect \cite{Alsing.et.al},  from the viewpoint of quantum information theory.
In the present paper, we extend the previous work in Ref.~\cite{HIUY}, which investigated
the entanglement-based description of the Minkowski vacuum state for a scalar field in a unified manner by connecting the
quantum states in the Rindler spacetime and the Kasner spacetime to the Dirac field.  
Our results show that modes associated with the Minkowski vacuum entanglement between the R and L regions are analytically connected to those in the F region and the P region. The results also indicate the time-like entanglement between the F region and the P region. In particular, we show these relations in an explicit manner, and it enables us to obtain fundamental quantities such as the thermal spectrum 
in all (F,P,R,L) regions covering the entire Minkowski spacetime. As far as we know, such an explicit demonstration of the analytic continuation-property 
of the general 4-dimensional spinor field has not been achieved so far. 

The rest of the paper is organized as follows. In Sec.~II the
massive Dirac field is quantized in the R region, the L region, the F region and the P region, where the explicit forms of the mode functions are given. 
In Sec.~III 
it is demonstrated that the solutions to the Dirac equation in the F region in terms of the Hankel function of the second kind are indeed the positive-frequency modes for the Minkowski vacuum state. The analytic continuation of the mode functions from the F region to the R and L regions is presented, together with that from the P region to the R and L regions. 
Subsequently, the relations of the mode functions between the F region, the R region, the L region and the P region are obtained. 
In Sec.~IV, using the Bogoliubov transformation between the two sets of the modes in the F region, we find the entanglement-based description of the Minkowski vacuum state. 
Combining the results of Sec.~III, we find the entanglement-based description of the Minkowski vacuum state in the R region, the L region, the F region and the P region in a unified manner. 
Sec.~V is devoted to a summary and conclusions. 
In Appendix~A, we present the matrices to transform the spinor between local Lorentz frames. In Appendix~B  verification of an ansatz for the
Minkowski vacuum state adopted in Sec.~IV is presented. 
In Appendix~C, a summary of equivalent mode functions
is presented. 

\section{Analysis of Solutions of Massive Dirac Fields in The R,L,F,P-Regions}
\label{Analysis of Spinors of Four-Dimensional Massive Dirac Fields in The R,L,F,P-Regions}

\begin{figure}[t]
  \begin{center}
   \includegraphics[width=110mm]{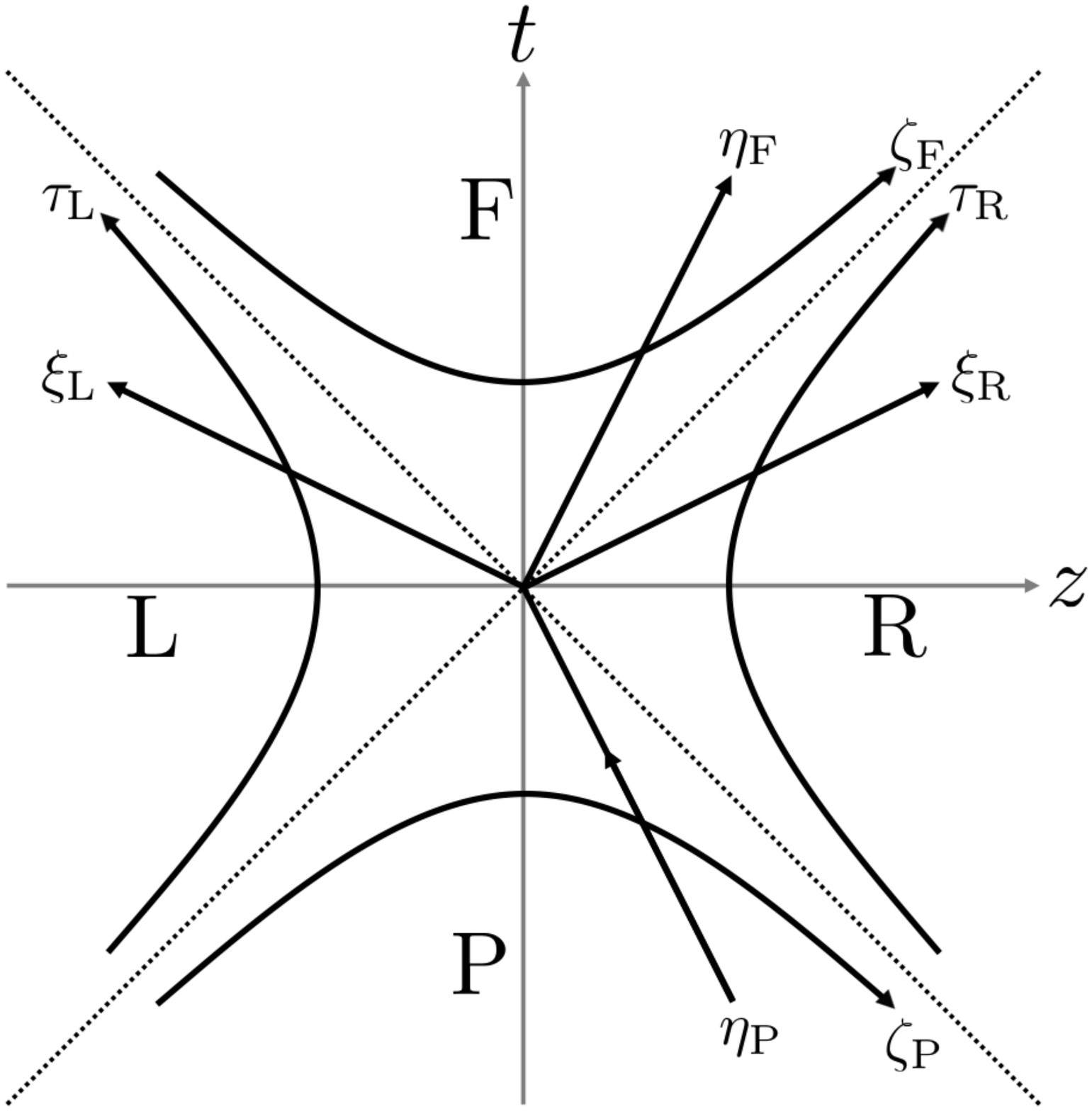}
  \end{center}
  \caption{Minkowski spacetime and the coordinates for the R region, the L region, the F region, 
	and the P region.}
 \label{Mcoordinate}
 \end{figure}
 
In this section, we derive the Dirac spinors from the Dirac equation defined in each region with the spin connection coefficient.
We consider a four-dimensional massive Dirac field 
in the four coordinate systems covering the entire Minkowski spacetime. 
The action of the massive Dirac field with a mass $m$ is given by 
\begin{equation}
S=\int d^4x \sqrt{-g} ~\bar\psi(i\gamma^\mu\nabla_\mu-m)\psi,
\end{equation}
where $\gamma^\mu$  are the gamma matrices in the curved spacetime. 
Here $\psi$ is a Dirac 4-spinor written in Dirac representation, 
which satisfies 
\begin{eqnarray}
\biggl[
i \gamma^\mu ({\partial \over \partial x^\mu}+\Gamma_\mu)-m
\biggr] \psi=0,
\label{DiracEQ}
\end{eqnarray}
where the spin connection coefficient is written as \begin{eqnarray}
\Gamma_\mu ={1\over 4} \gamma_\nu \biggl(
{\partial \gamma^\nu \over \partial x^\mu}+\Gamma^\nu_{\lambda \mu} \gamma^\lambda
\biggr),
\label{Scoef}
\end{eqnarray} and Dirac gamma matrices satisfy the Clifford algebra $\bigl\{ \gamma^\mu ,\gamma^\nu \bigr\}=2g^{\mu\nu}$. 
In this section, we follow the basic procedure to derive Dirac spinors in curved spacetimes that can be found in many literatures (see Refs.~\cite{Vanzella,D.Roy} for the derivation of the spinors). 
In Sec. \ref{AppendixE}, we show a different way to derive Dirac spinor in each region from Minkowski mode.
Minkowski spacetime is described by the global coordinate $(t,x,y,z)$ with the line element
\begin{eqnarray}
ds^2=dt^2-dx^2-dy^2-dz^2.
\end{eqnarray}
The right Rindler region (R region) is described by the coordinates
($\tau_{\rm R},\xi_{\rm R}$), which are related to the Minkowski coordinates $(t,z)$ by
\begin{equation}
t={1\over a}e^{a\xi_{\rm R}}\sinh a\tau_{\rm R},~~
z={1\over a}e^{a\xi_{\rm R}}\cosh a\tau_{\rm R}.
\label{Rregion}
\end{equation}
The left Rindler region (L region) is described by the coordinates ($\tau_{\rm L},\xi_{\rm L}$) 
defined as
\begin{equation}
t={1\over a}e^{a\xi_{\rm L}}\sinh a\tau_{\rm L},~~
z=-{1\over a}e^{a\xi_{\rm L}}\cosh a\tau_{\rm L}.
\label{Lregion}
\end{equation}
The future Kasner region (F region) is described by the expanding-Kasner spacetime
($\eta_{\rm F},\zeta_{\rm F}$) defined as
\begin{equation}
t={1\over a}e^{a\eta_{\rm F}}\cosh a\zeta_{\rm F},~~
z={1\over a}e^{a\eta_{\rm F}}\sinh a\zeta_{\rm F}.
\label{Fregion}
\end{equation}
The past Kasner region (P region) is described by the shrinking-Kasner spacetime ($\eta_{\rm P},\zeta_{\rm P}$) defined as
\begin{equation}
t=-{1\over a}e^{-a\eta_{\rm P}}\cosh a\zeta_{\rm P},~~
z={1\over a}e^{-a\eta_{\rm P}}\sinh a\zeta_{\rm P}.
\label{Pregion}
\end{equation}
Each coordinate system covers a quarter of the Minkowski spacetime, as shown in Fig.~\ref{Mcoordinate}.

\vspace{4mm}
    \begin{center}
    \begin{tabular}{ccc}
      \hline
      ~~~~~~~~region~~~~~~  & ~~~~~~variable~1~~~~~  & ~~~~~~variable~2~~~~~~  \\
      \hline\hline\addlinespace[2pt]
      \vspace{1mm} R$\longleftrightarrow$F & $\displaystyle {\tau_{\rm R}=\zeta_{\rm F}-{\pi i\over 2a}}$  & $\displaystyle{\xi_{\rm R}=\eta_{\rm F}+{\pi i\over 2a}}$ \\
      \vspace{1mm} R$\longleftrightarrow$P &  $\displaystyle {\tau_{\rm R}=-\zeta_{\rm P}-{\pi i\over 2a}}$  &  $\displaystyle{\xi_{\rm R}= -\eta_{\rm P}-{\pi i\over 2a}}$ \\
      \vspace{1mm} R$\longleftrightarrow$L & $\displaystyle {\tau_{\rm R}=-\tau_{\rm L}-{\pi i\over a}}$  & $\displaystyle{\xi_{\rm R}=\xi_{\rm L}}$ \\
      \vspace{1mm} F$\longleftrightarrow$L & $\displaystyle {\zeta_{\rm F}=-\tau_{\rm L}-{\pi i\over 2a}}$  & $\displaystyle{\xi_{\rm L}=\eta_{\rm F}+{\pi i\over 2a}}$ \\
      P$\longleftrightarrow$L & $\displaystyle {\zeta_{\rm P}=\tau_{\rm L}+{\pi i\over 2a}}$  & $\displaystyle{\eta_{\rm P}=-\xi_{\rm L}-{\pi i\over 2a}}$  \vspace{1mm}\\
      \hline
            \label{Analytic continuation of valuables}
    \end{tabular}
    \vspace{-4mm}
      \tblcaption{Analytic continuation of variables 
      }
  \end{center}
\subsection{Dirac Fied in R,L-Rindler Region}
\label{Dirac Field in R(L)-Region}
We start from a brief review of the solution of the Dirac equation in the R,L region where the line element is 
\begin{eqnarray}
ds^2=e^{2a\xi_{\Lambda}}d\tau_{\Lambda}^2-dx^2-dy^2-e^{2a\xi_{\Lambda}}d\xi_{\Lambda}^2,~~(\Lambda={\rm R~or~L}).
\end{eqnarray}
Here, $(\tau_{\rm R},\xi_{\rm R})$ are the right Rindler coordinates, while  $(\tau_{\rm L},\xi_{\rm L})$ is the Left Rindler coordinate, defined in Eqs.~(\ref{Rregion}) and (\ref{Lregion}). Also, note that we use Greek letter Lambda $\Lambda={\rm R,L}$ to denote R region or L region. We use it to denote the expression which is valid for both of R and L region.
Also, we use the notation $\gamma^\mu_{\Lambda}$ to denote the Dirac matrices in R, L Rindler region, and $\gamma^\mu$ for Dirac matrices in flat spacetime, respectively. 
Using the property of the gamma matrix in Minkowski spacetime, $(\gamma^0)^2=1$ and $(\gamma^1)^2=(\gamma^2)^2=(\gamma^3)^2=-1$, we  have the relation between $\gamma^\mu_{\Lambda}$ and $\gamma^\mu$:
\begin{eqnarray}
\gamma^0_{\Lambda} = e^{-a\xi_{\Lambda}} \gamma^0~,~~ \gamma^1_{\Lambda} = \gamma^1~,~~ 
\gamma^2_{\Lambda} = \gamma^2~,~~ \gamma^3_{\Lambda} =e^{-a\xi_{\Lambda}} \gamma^3 .\nonumber
\end{eqnarray}
The spin connection coefficient Eq.~(\ref{Scoef}) in R,L region is derived as
\begin{eqnarray}
\Gamma_\mu = {\rm diag.} ({a\over 2} \gamma^0 \gamma^3,0,0,0).
\end{eqnarray}
By adopting the notation $\gamma^0 \gamma^j=\alpha_j,\gamma^0=\beta$, 
the Dirac equation on the R,L region reduces to
\begin{align}
\biggl[
i{\partial \over \partial \tau_{\Lambda}} +ie^{a\xi_{\Lambda}}(\alpha_1{\partial\over\partial x}+\alpha_2{\partial \over \partial y})+i\alpha_3{\partial \over \partial \xi_{\Lambda}}
+i{a\over 2}\alpha_3 -e^{a\xi_{\Lambda}} \beta m
\biggr] \psi_{\omega, {\bm k}_\perp}^{\Lambda,n}=0.
\label{Drl}
\end{align}
Explicit forms of the matrices in Dirac equation with Dirac representation take following forms:
\begin{eqnarray}
\alpha_1=\begin{pmatrix}
0&0&0&1 \\
0&0&1&0\\
0&1&0&0\\
1&0&0&0\\
\end{pmatrix} \nonumber ,~
\alpha_2=\begin{pmatrix}
0&0&0&-i \\
0&0&i&0\\
0&-i&0&0\\
i&0&0&0
\end{pmatrix}\nonumber ,~
\alpha_3=\begin{pmatrix}
0&0&1&0 \\
0&0&0&-1\\
1&0&0&0\\
0&-1&0&0
\end{pmatrix}\nonumber ,~
\beta=\begin{pmatrix}
1&0&0&0 \\
0&1&0&0\\
0&0&-1&0\\
0&0&0&-1
\end{pmatrix}.
\end{eqnarray}
We set up the ansatz for four spinors of the Dirac equation in Rindler spacetime as 
\begin{eqnarray}
\psi_{\omega, {\bm k}_\perp}^{\Lambda,n}(\tau_{\Lambda},\xi_{\Lambda},\bm{x}_\perp)\equiv f_{\omega,{\bm k}_\perp}^{\Lambda,n}(\xi_{\Lambda}) e^{i{{\bm k}_\perp}\cdot{\bm x_\perp}} e^{-i\omega \tau_{\Lambda}},
\label{RmodeED}
\end{eqnarray}
where $n={\rm u}$ denotes the up spin while $n={\rm d}$ denotes down spin. The mode expansion of the Dirac field in the R,L region is written as
\begin{eqnarray}
&&{\hat \Psi}(\tau_{\Lambda},\xi_{\Lambda},\bm{x}_\perp)
=\sum_{ n={\rm u,d}} \int_0 ^{\infty} d\omega \int^\infty_{-\infty} d^2{\bm k}_\perp \biggl( 
{\hat c^{\Lambda,  n}_{\omega, {\bm k}_\perp} \psi_{\omega, {\bm k}_\perp}^{\Lambda,n}(\tau_{\Lambda},\xi_{\Lambda},\bm{x}_\perp)}+
{\hat d_{\omega, {\bm k}_\perp}^{{\Lambda},n\dagger} (\psi_{\omega, {\bm k}_\perp}^{\Lambda,n}(\tau_{\Lambda},\xi_{\Lambda},\bm{x}_\perp))^{\rm C}}
\biggr),
\label{RmodeE}
\end{eqnarray}
where $\psi^{\rm C}$ is the spinor which is corresponding to the anti-particle, obtained by charge conjugation of $\psi$ defined by $\psi^{\rm C}=i\gamma^2\psi^*$.
We integrate the positive range of $\omega$, since $\omega$ is energy in the Rindler region as is shown by the ansatz (\ref{RmodeED}) apparently, i.e. we focus on the positive frequency mode.
We require the creation and annihilation operators of the Dirac fermion and anti-fermion to satisfy the following anti-commutation relations:
\begin{equation}
\{ {\hat \Omega^{\Lambda,n}_{\omega, {\bm k}_\perp}} , {\hat \Omega^{\dagger{\Lambda,n'}}_{\omega', {\bm k'}_\perp}} \}=\delta(\omega-\omega')\delta({\bm k}_\perp-{\bm k}_\perp')\delta_{ n n'},
\end{equation}
with all other anti-commutators vanishing.
Here, $\hat\Omega^{\Lambda,n}_{\omega,{\bm k}_\perp}$ is operators like $\hat c^{\Lambda,n}_{\omega,{\bm k}_\perp}$ or $\hat d^{\Lambda,n}_{\omega,{\bm k}_\perp}$ in Eq.~(\ref{RmodeE}). The normalization condition of mode functions takes the following form with the Dirac inner product,
\begin{eqnarray}
 (\psi^{\Lambda,n}_{\omega,{\bm k}_\perp},\psi^{\Lambda,n'}_{\omega',\bm k'_\perp})_{\rm D}=\int_{\Sigma} d\Sigma_\mu
  \overline{\psi^{\Lambda,n}_{\omega,{\bm k}_\perp} }\gamma^\mu_{\rm \Lambda}
  \psi^{\Lambda,n'}_{\omega',{\bm k'_\perp}}=\delta(\omega-\omega')\delta({\bm k}_\perp-{\bm k'_\perp})\delta_{ n n'},
\label{Rnormal}
\end{eqnarray}
where we use the overline to denote Dirac conjugate, $\overline{\psi^{\Lambda,n}_{\omega,{\bm k}_\perp} }\equiv (\psi^{\Lambda,n}_{\omega,{\bm k}_\perp})^\dagger \gamma^0$.
Here $(\cdot,\cdot)_{\rm D}$ denotes the Dirac inner product.
We define $ d\Sigma_\mu=d\Sigma n_\mu$, where 
$d\Sigma$ denotes the 3D spatial volume element on the Cauchy hypersurface 
$\Sigma$ to determine the time slice where the normalization condition of the spinor is defined, and $n_\mu$ is the normal vector vertical to $\Sigma$.
Plugging the ansatz Eq.~(\ref{RmodeED}) into the Dirac equation yields
\begin{eqnarray}
\omega f_{\omega,{\bm k}_\perp}^{\Lambda,n}(\xi_{\rm \Lambda})
=\biggl[\gamma^0 m e^{a\xi_{\rm \Lambda}} -i{a \over 2}\alpha_3 -i\alpha_3 {\partial\over\partial\xi_{\rm \Lambda}}+e^{a\xi_{\rm \Lambda}} (\alpha_1 k_1+ \alpha_2 k_2) \biggr]f_{\omega,{\bm k}_\perp}^{\Lambda,n}(\xi_{\Lambda}).
\label{fDirac}
\end{eqnarray}
In order to describe Dirac equation in more detail, we define the solution as 2 spinors which takes the form
\begin{eqnarray}
f_{\omega,{\bm k}_\perp}^{\Lambda,n}(\xi_{\rm \Lambda})=
\left.
\begin{pmatrix}
\chi_{1}^{\Lambda,n} (\xi_{\rm \Lambda}) \\
\chi_{2}^{\Lambda,n} (\xi_{\rm \Lambda})
\end{pmatrix}
\right..
\label{definition-of-f}
\end{eqnarray}
In terms of the two spinor, the Dirac equation is described as the following two equations:
\begin{eqnarray}
 \omega \chi_{1}^{\Lambda,n}&=& me^{a\xi_{\rm \Lambda}} \chi_{1}^{\Lambda,n}-i{a\over 2} \sigma^3\chi_{2,n}^{\rm \Lambda}
- i \sigma^3 {\partial \chi_{2}^{\Lambda,n}\over \partial \xi_{\Lambda}}+e^{a\xi} k_1 \sigma^1 \chi_{2}^{\Lambda,n}+e^{a\xi_{\Lambda}} k_2 \sigma^2 \chi_{2}^{\Lambda,n}, \nonumber \\
 \omega \chi_{2}^{\Lambda,n}&=& -me^{a\xi_{\Lambda}} \chi_{2}^{\Lambda,n}-i{a\over 2} \sigma^3\chi_{1}^{\Lambda,n} 
- i \sigma^3{\partial \chi_{1}^{\Lambda,n}\over \partial \xi_{\Lambda}}+e^{a\xi_{\Lambda}} k_1 \sigma^1 \chi_{1}^{\Lambda,n}+e^{a\xi_{\Lambda}} k_2 \sigma^2 \chi_{1}^{\Lambda,n}.
\nonumber
\end{eqnarray}
These equations reduce to the same equation when we take the massless limit of the Dirac equation, resulting in the Weyl equation. 
After combining these equations, we obtain the following second order differential equations:
\begin{eqnarray}
\displaystyle {1\over a^2}{\partial^2\over\partial\xi^2_{\Lambda}}\chi_{1\{2\}}^{\Lambda,n}=\biggl[{\kappa^2\over a^2}e^{2a\xi_{\Lambda}}+{1\over4}-{\omega^2 \over a^2}\biggr]\chi_{1\{2\}}^{\Lambda,n}-{i\omega \over a} \sigma^3\chi_{2\{1\}}^{\Lambda,n},
\label{Simultaneous1}
\end{eqnarray}
where $\kappa$ is the effective mass: $\kappa=\sqrt{m^2+{\bm k}_\perp^2}$. Note that this corresponds to Eqs. (3.92) and (3.93) in Ref.~\cite{Higuchi}.
The braket $\{\}$ after the suffix implies that the form of Eq.~(\ref{Simultaneous1}) is unchanged under the swap of suffix 1 and 2, but $\chi_1$ and $\chi_2$ are still mixed. In order to obtain independent equations, we define
\begin{eqnarray}
\phi^{\Lambda,n}_{\pm}(\xi_{\Lambda})= \chi^{\Lambda,n}_{1}(\xi_{\rm \Lambda})\mp \chi^{\Lambda,n}_{2}(\xi_{\rm \Lambda})=
\begin{pmatrix}
\varrho^{\Lambda,n}_{\pm} (\xi_{\rm \Lambda}) \\
\varsigma^{\Lambda,n}_{\pm} (\xi_{\rm \Lambda})
\end{pmatrix}.
\label{definition-of-phi}
\end{eqnarray}
The spinors $\phi^{\Lambda,n}_{\pm}(\xi_{\rm \Lambda})$ corresponds to those in the Weyl representation. 
This reflects the general relation between the Dirac matrices in the Dirac and Weyl  representations: $\psi_{\rm Weyl}=U\psi_{\rm Dirac}$, where 
\begin{eqnarray}
U={1\over\sqrt{2}}
\begin{pmatrix}
1&1\\
1&-1
\end{pmatrix}.
\nonumber
\end{eqnarray}
We can reproduce the form of Dirac spinor in the Dirac representation as  
\begin{eqnarray}
f^{\Lambda,n}_{\omega,{\bm k}_\perp}={1\over2}
\begin{pmatrix}
\varrho^{\Lambda,n}_{+} (\xi_{\rm \Lambda})+\varrho^{\Lambda,n}_{-} (\xi_\Lambda) \\
\varsigma^{\Lambda,n}_{+} (\xi_{\rm \Lambda})+\varsigma^{\Lambda,n}_{-} (\xi_\Lambda)\\
-\varrho^{\Lambda,n}_{+} (\xi_{\rm \Lambda}) +\varrho^{\Lambda,n}_{-} (\xi_\Lambda) \\
-\varsigma^{\Lambda,n}_{+} (\xi_{\rm \Lambda})+\varsigma^{\Lambda,n}_{-} (\xi_\Lambda)\\
\end{pmatrix},
\label{Drep}
\end{eqnarray}
by plugging definitions Eq.~(\ref{definition-of-phi}) and Eq.~(\ref{definition-of-f}).
Equation~(\ref{Simultaneous1}) written in terms of $\phi^{\Lambda,n}_{\pm}(\xi_\Lambda)$ is
\begin{eqnarray}
{1\over a^2}{\partial^2\over\partial\xi_{\rm \Lambda}^2}\phi^{\Lambda,n}_{\pm}(\xi_\Lambda)=\biggl[{\kappa^2\over a^2}e^{2a\xi_\Lambda}+{1\over4}-{\omega^2 \over a^2}\biggr]\phi^{\Lambda,n}_{\pm}(\xi_{\rm \Lambda}) \pm{i\omega \over a}\sigma^3\phi^{\Lambda,n}_{\pm} (\xi_\Lambda).
\label{2spiBessel}
\end{eqnarray}
Thus, we obtain the following Bessel-type differential equations:
\begin{eqnarray}
{1\over a^2}{\partial^2\over\partial\xi_\Lambda^2} \varrho^{\Lambda,n}_{\pm} (\xi_\Lambda)
=\biggl[{\kappa^2\over a^2}e^{2a\xi_\Lambda}+\biggl({i\omega \over a}\pm{1\over2}\biggr)^2\biggr] \varrho^{\Lambda,n}_{\pm} (\xi_\Lambda),  \nonumber\\
{1\over a^2}{\partial^2\over\partial\xi_\Lambda^2}\varsigma^{\Lambda,n}_{\pm} (\xi_\Lambda)
=\biggl[{\kappa^2\over a^2}e^{2a\xi_\Lambda}+\biggl({i\omega \over a}\mp{1\over2}\biggr)^2\biggr]
\varsigma^{\Lambda,n}_{\pm} (\xi_{\Lambda}). \nonumber
\end{eqnarray}
Solutions of these differential equations are written in terms of modified Bessel functions $K_{i\omega/a\pm1/2}({\kappa\over a} e^{a\xi_\Lambda})$, $I_{i\omega/a\pm1/2}({\kappa\over a} e^{a\xi_\Lambda})$.
The solution written in terms of $I_{i\omega/a\pm1/2}({\kappa\over a} e^{a\xi_\Lambda})$ 
cannot be normalized since diverges in the limit $\xi_\Lambda\to\infty$. Therefore, we need to choose the following solutions:
\begin{equation}
\label{define-c-coeffs}
\begin{split}
&\varrho^{\Lambda,n}_+(\xi_\Lambda)=c^{n(\Lambda)}_1 K_{i\omega/a+1/2} (q_\Lambda)~,~~
\varrho^{\Lambda,n}_-(\xi_\Lambda)=c^{n(\Lambda)}_2 K_{i\omega/a-1/2} (q_\Lambda)~,\\
&\varsigma^{\Lambda,n}_+(\xi_\Lambda)=c^{n(\Lambda)}_3 K_{i\omega/a-1/2} (q_\Lambda)~,~~
\varsigma^{\Lambda,n}_-(\xi_\Lambda)=c^{n(\Lambda)}_4K_{i\omega/a+1/2} (q_\Lambda)~,
\end{split}
\end{equation}
where 
\begin{equation}
q_{\rm R}= \frac{\kappa}{a}e^{a\xi_{\rm R}},~ 
q_{\rm L}= \frac{\kappa}{a}e^{a\xi_{\rm L}},
\label{qRqL}
\end{equation}
as solutions of quantum field theory.  
Here, we add the coefficients to each solutions since Bessel differential equations after Eq.~(\ref{2spiBessel}) are independent. These coefficients have information of the spin direction and are also constrained by the normalization condition. By plugging Eq.~(\ref{Drep}) with the four solutions defined by Eq.~(\ref{define-c-coeffs}) into the Dirac equation (\ref{fDirac}), one finds
\begin{eqnarray}
-c^{n(\Lambda)}_1 m -c^{n(\Lambda)}_2 i\kappa-c^{n(\Lambda)}_4(k_1-ik_2)=0,\nonumber\\
c^{n(\Lambda)}_1(k_1+ik_2)-c^{n(\Lambda)}_3i\kappa-c^{n(\Lambda)}_4 m=0,\nonumber\\
c^{n(\Lambda)}_1 i\kappa-c^{n(\Lambda)}_2 m+c^{n(\Lambda)}_3(k_1-ik_2)=0,\nonumber\\
-c^{n(\Lambda)}_2(k_1+ik_2)-c^{n(\Lambda)}_3 m+c^{n(\Lambda)}_4 i\kappa=0.
\nonumber
\end{eqnarray}
If ${\bm k}_\perp \neq 0$, then the first two equations are equivalent to the last two.  Thus, there are two 
linearly independent solutions for the vector $(c^{n(\Lambda)}_1,c^{n(\Lambda)}_2,c^{n(\Lambda)}_3,c^{n(\Lambda)}_4)$.  Two linearly independent solutions can be
chosen as
\begin{equation}
(c^{{\rm u}(\Lambda)}_1,c^{{\rm u}(\Lambda)}_2,c^{{\rm u}(\Lambda)}_3,c^{{\rm u}(\Lambda)}_4) = (1, im/\kappa, -i(k_1 + ik_2)/\kappa,0),
\end{equation}
and
\begin{equation}
(c^{{\rm d}(\Lambda)}_1,c^{{\rm d}(\Lambda)}_2,c^{{\rm d}(\Lambda)}_3,c^{{\rm d}(\Lambda)}_4) = (0, (k_1-ik_2)/\kappa, m/\kappa,-i).
\end{equation}
This choice gives the solutions in Eq.~(\ref{SR+}), which are confirmed to be orthogonal.  One can also choose
\begin{equation}
(c^{{\rm u}(\Lambda)}_1,c^{{\rm u}(\Lambda)}_2,c^{{\rm u}(\Lambda)}_3,c^{{\rm u}(\Lambda)}_4) = (m/\kappa,i,0,(k_1+ik_2)/\kappa),
\end{equation}
and
\begin{equation}
(c^{{\rm d}(\Lambda)}_1,c^{{\rm d}(\Lambda)}_2,c^{{\rm d}(\Lambda)}_3,c^{{\rm d}(\Lambda)}_4)=(-(k_1-ik_2)/\kappa,0,i,m/\kappa).
\end{equation}
This choice gives the solutions in Eq.~(\ref{SR+2}), which are orthogonal.  
These solutions are valid also for ${\bm k}_\perp =0$.
The other solutions can be obtained
in a similar manner.
The coefficients  $\{c^{n(\Lambda)}_1,c^{n(\Lambda)}_2,c^{n(\Lambda)}_3,c^{n(\Lambda)}_4\}$  are not arbitrary since the functions in Eq.~(\ref{Drep}) have to satisfy the Dirac equation (\ref{Drl}) though each of solutions in Eq.~(\ref{define-c-coeffs}) are derived from four independent Bessel type differential equations. In addition, when we have the coefficients  $\{c^{n(\Lambda)}_1,c^{n(\Lambda)}_2,c^{n(\Lambda)}_3,c^{n(\Lambda)}_4\}$, the other set of coefficients  $\{c^{\overline n(\Lambda)}_1,c^{\overline n(\Lambda)}_2,c^{\overline n(\Lambda)}_3,c^{\overline n(\Lambda)}_4\}$  can be obtained after taking the charge conjugation of the spinor, here the overline on the index {\it n} denotes the swap of the spin up and down.

Solutions of the Dirac equation in four-dimensional Rindler region have been found previously, and was given in other paper (see, e.g., Refs.~\cite{SuzukiYamada,oriti}). 
They can be reproduced with certain choices of the coefficients in Eq.~(\ref{Drep}).
Here, we further simplify the solutions in R region by the following form:
\begin{eqnarray}
f^{{\rm R}, n}_{\omega,{\bm k}_\perp}(\xi_{\rm \Lambda})= A_{\omega,{\bm k}_\perp}
S^{{\rm R}, n}_{\omega,{\bm k}_\perp}(\xi_{\rm \Lambda}).
\end{eqnarray}
Here $A_{\omega,{\bm k}_\perp}$ are the real and positive normalization constants determined by the orthonormal condition (\ref{Rnormal}), 
and $S^{{\rm R}, n}_{\omega,{\bm k}_\perp}(\xi_{\rm \Lambda})$ are linearly independent spinors which have internal degrees of freedom corresponding to the spin direction.

Two linearly independent and orthogonal spinors can be chosen as follows:
\begin{align}
S_{\omega,{\bm k}_\perp}^{{\rm R},n}
(\xi_{\rm R})=
s_{{\bm k}_\perp}^{(+,n)}K_{i\omega/a+1/2}(q_{\rm R})+s_{{\bm k}_\perp}^{(-,n)}K_{i\omega/a-1/2} (q_{\rm R}),
\label{SR+}
\end{align}
where we define
\begin{align}
s_{{\bm k}_\perp}^{(+,{\rm u})}=
\begin{pmatrix}
1\\
0\\
-1\\
0\\
\end{pmatrix},~~
s_{{\bm k}_\perp}^{(-,{\rm u})}
={i\over\kappa}
\begin{pmatrix}
m\\
-(k_1+ik_2)\\
m\\
k_1+ik_2\\
\end{pmatrix}, ~~
s_{{\bm k}_\perp}^{(+,{\rm d})}=-i
\begin{pmatrix}
0\\
1\\
0\\
1\\
\end{pmatrix}, ~~
s_{{\bm k}_\perp}^{(-,{\rm d})}={1\over\kappa}
\begin{pmatrix}
k_1-ik_2\\
m\\
k_1-ik_2\\
-m\\
\end{pmatrix}.
\label{sdef}
\end{align}
Then, the normalization constant is
\begin{eqnarray}
A_{\omega,{\bm k}_\perp}={1\over 2\pi}\sqrt{\kappa \cosh{\pi\omega/ a}\over 2\pi^2 a}.
\label{RN}
\end{eqnarray}
The choice of the base of mode expansion Eq.~(\ref{RmodeE}) is arbitrary, though we here assign spinors (\ref{SR+}) as the solution of the R region for simplicity.
On the other hand, we assign the other choice of two linearly independent and orthogonal spinors
\begin{align}
S_{\omega,{\bm k}_\perp}^{{\rm L},n}
(\xi_{\rm L})&=\tilde{s}_{{\bm k}_\perp}^{(+,n)}K_{i\omega/a+1/2}(q_{\rm L})+\tilde{s}_{{\bm k}_\perp}^{(-,n)}K_{i\omega/a-1/2} (q_{\rm L})
\label{SR+2}
\end{align}
for the solution in L region. 
Here, we define
\begin{align}
\tilde{s}_{{\bm k}_\perp}^{(+,{\rm u})}=
{1\over\kappa}
\begin{pmatrix}
m\\
k_1+ik_2\\
-m\\
k_1+ik_2\\
\end{pmatrix},~~
\tilde{s}_{{\bm k}_\perp}^{(-,{\rm u})}
=i
\begin{pmatrix}
1\\
0\\
1\\
0\\
\end{pmatrix}, ~~
\tilde{s}_{{\bm k}_\perp}^{(+,{\rm d})}={1\over\kappa}
\begin{pmatrix}
-(k_1-ik_2)\\
m\\
k_1-ik_2\\
m\\
\end{pmatrix}, ~~
\tilde{s}_{{\bm k}_\perp}^{(-,{\rm d})}=i
\begin{pmatrix}
0\\
1\\
0\\
-1\\
\end{pmatrix}. 
\end{align}
We write the corresponding spinor mode functions as
\begin{eqnarray}
&&\psi_{\omega, {\bm k}_\perp}^{{\rm L},n}(\tau_{\rm L},\xi_{\rm L},\bm{x}_\perp)\equiv f_{\omega,{\bm k}_\perp}^{{\rm L},n}(\xi_{\rm L}) e^{i{{\bm k}_\perp}\cdot{\bm x_\perp}} e^{-i\omega \tau_{\rm L}},\\
&&f^{{\rm L},n}_{\omega,{\bm k}_\perp}(\xi_{\rm L})
= A_{\omega,{\bm k}_\perp}
S^{{\rm L}, n}_{\omega,{\bm k}_\perp}(\xi_{\rm L}).
\end{eqnarray}
The normalization constant is the same as before Eq.~(\ref{RN}), and sets of the spinor are orthogonal,
$(S_{\omega,{\bm k}_\perp}^{{\rm R},n},S_{\omega,{\bm k}_\perp}^{{\rm R},\overline{n}})_{\rm D}=(S_{\omega,{\bm k}_\perp}^{{\rm L},n}, S_{\omega,{\bm k}_\perp}^{{\rm L},\overline{n}})_{\rm D}=0$.
The form of the spinor are related by the following linear equations:
\begin{eqnarray}
S_{\omega,{\bm k}_\perp}^{\rm L,u} (\xi_{\rm \Lambda})
&=&{m\over\kappa} 
S_{\omega,{\bm k}_\perp}^{\rm R,u}
(\xi_{\rm \Lambda})
+ i{{k_1+ik_2}\over\kappa}
S_{\omega,{\bm k}_\perp}^{\rm R,d}(\xi_{\rm \Lambda}), \\
S_{\omega,{\bm k}_\perp}^{\rm L,d}
(\xi_{\rm \Lambda})&=&-{{k_1-ik_2}\over\kappa} 
S_{\omega,{\bm k}_\perp}^{\rm R,u}
(\xi_{\rm \Lambda})+
i{m\over \kappa} S_{\omega,{\bm k}_\perp}^{\rm R,d}
(\xi_{\rm \Lambda}),
\end{eqnarray}
thus, we can use the spinor form $\psi^{\rm L}$ as an base of the solution in R region actually. 
We can describe any solutions in 4 dimensional Rindler region with the combination of bases of the mode, and we can realize any 2 or 3 dimensional solution by setting $k_{j}=0~(j=1,2)$ appropriately. 

\subsection{Dirac Field in the F(P)-Kasner Region}
\label{Dirac Field in F-Kasner Region}
In this subsection, we derive Dirac spinors in F and P region covered by Kasner coordinates. The line element of the F(P) region is
\begin{align}
&{\rm F~region~:~}ds^2=e^{2a\eta_{\rm F}}d\eta_{\rm F}^2-dx^2-dy^2-e^{2a\eta_{\rm F}}d\zeta_{\rm F}^2,\\
&{\rm P~region~:~}ds^2=e^{-2a\eta_{\rm P}}d\eta_{\rm P}^2-dx^2-dy^2-e^{-2a\eta_{\rm P}}d\zeta_{\rm P}^2,
\end{align}
with local coordinates defined by Eqs.~(\ref{Fregion}) and (\ref{Pregion}).
Using the property of the gamma matrices in Minkowski spacetime
, $(\gamma^0)^2=1$ and $(\gamma^1)^2=(\gamma^2)^2=(\gamma^3)^2=-1$, we have the gamma matrices in the F(P) region $\gamma^\mu_{\rm F(P)}$ as
\begin{eqnarray}
&&{\rm F~region~:~}\gamma^0_{\rm F} = e^{-a\eta_{\rm F}} \gamma^0~,~~ \gamma^1_{\rm F} = \gamma^1~,~~ 
\gamma^2_{\rm F} = \gamma^2~,~~ \gamma^3_{\rm F} =e^{-a\eta_{\rm F}} \gamma^3~,\\
&&{\rm P~region~:~}\gamma^0_{\rm P} = e^{a\eta_{\rm P}}  \gamma^0,~ \gamma^1_{\rm P} = \gamma^1~,~~
\gamma^2_{\rm P} = \gamma^2,~ \gamma^3_{\rm P} =e^{a\eta_{\rm P}} \gamma^3~.
\end{eqnarray}
The spin connection is given as
\begin{align}
&{\rm F~region~:~}\Gamma_\mu = {\rm diag}.(0,0,0,{a\over 2} \gamma^0 \gamma^3)~,\\
&{\rm P~region~:~}\Gamma_\mu = {\rm diag}.(0,0,0,-{a\over 2} \gamma^0 \gamma^3)~.
\end{align}
The Dirac equation takes the following form:
\begin{align}
&{\rm F~region~:~}
\biggl[
i{\partial \over \partial \eta_{\rm F}} +ie^{a\eta_{\rm F}}(\alpha_1{\partial\over\partial x}+\alpha_2{\partial \over \partial y})+i\alpha_3{\partial \over \partial \zeta_{\rm F}}
+i{a\over 2} -e^{a\eta_{\rm F}} \beta m
\biggr]
\psi^{\Xi,n({\rm F})}_{\omega, {\bm k}_\perp}=0,\\
&{\rm P~region~:~}
\biggl[
i{\partial \over \partial \eta_{\rm P}} +ie^{-a\eta_{\rm P}}(\alpha_1{\partial\over\partial x}+\alpha_2{\partial \over \partial y})+i\alpha_3{\partial \over \partial \zeta_{\rm P}}
-i{a\over 2}-e^{-a\eta_{\rm P}} \beta m
\biggr]\psi^{\Xi,n({\rm P})}_{\omega, {\bm k}_\perp}=0.
\label{DDD}
\end{align}
We describe the ansatz for four spinors of the Dirac equation in Kasner spacetime as
\begin{eqnarray}
&&\psi^{\Xi,n(\Theta)}_{\omega, {\bm k}_\perp}(\eta_\Theta,\zeta_\Theta,{\bm x}_\perp) \equiv 
f^{\Xi,n(\Theta)}_{\omega, {\bm k}_\perp}(\eta_\Theta) e^{i{{\bm k}_\perp}\cdot{\bm x_\perp}} e^{-i\omega \zeta_\Theta},
\label{anzatzF}
\end{eqnarray}
where $\Theta$ means ``F" or ``P". 
We distinguish two different orthonormal bases of the solution by the label ``$\Xi$" which denote ``${\rm M}"$ or ``${\rm K}$".
The label ``M" denotes the Minkowski mode, which is clearly expressed with Hankel functions  $H^{(1)}_\nu, H^{(2)}_\nu$. In contrast, the label ``K" denotes Kasner modes which is actually connected to the Rindler mode later, both of them certainly satisfy the Dirac equation in each region. We clarify the correspondence of modes in four regions in the latter part of this paper.
Also, ``$\Theta$" denote F region or P region just as ${\rm K}$ denoted R and L region in previous subsection. Later in this subsection, one can find that $\psi^{{\rm K},n(\Theta)}_{\omega, {\bm k}_\perp}$ is certainly written with Bessel function $J_\nu(x)$, while $\psi^{{\rm M},n(\Theta)}_{\omega, {\bm k}_\perp}$ is written with Hankel functions.
Then the Dirac field is expanded in this region as
\begin{align}
{\hat \Psi}(\eta_\Theta,\zeta_\Theta,{\bm x}_\perp)
=\sum_{ n={\rm u,d}} \int_{-\infty} ^{\infty} d\omega \int^\infty_{-\infty} d^2{\bm k}_\perp \biggl( 
{\hat a^{{\rm M},n(\Theta)}_{\omega, {\bm k}_\perp}} 
\psi^{{\rm M},n(\Theta)}_{\omega, {\bm k}_\perp}
+
{\hat b^{{\rm M},n(\Theta)\dagger}_{\omega, {\bm k}_\perp}
(\psi^{{\rm M},n(\Theta)}_{\omega, {\bm k}_\perp})^{\rm C}} \biggr)\nonumber \\
=\sum_{ n={\rm u,d}} \int_{-\infty} ^{\infty} d\omega \int^\infty_{-\infty} d^2{\bm k}_\perp \biggl( 
{\hat c^{{\rm K},n(\Theta)}_{\omega, {\bm k}_\perp} \psi^{{\rm K},n(\Theta)}_{\omega, {\bm k}_\perp}}+
{\hat d^{{\rm K},n(\Theta)\dagger}_{\omega, {\bm k}_\perp} (
\psi^{{\rm K},n(\Theta)}_{\omega, {\bm k}_\perp}
)^{\rm C}}
\biggr).
\label{ExpandF}
\end{align}
We require the creation and annihilation operators of the Dirac fermion and anti-fermion to satisfy the following anti-commutation relations:
\begin{equation}
\{ {\hat \Omega^{\Xi,n(\Theta)}_{\omega, {\bm k}_\perp}} , {\hat \Omega^{\Xi,n'(\Theta)\dagger}_{\omega', {\bm k'}_\perp}} \}=\delta(\omega-\omega')\delta({\bm k}_\perp-{\bm k}_\perp')\delta_{ n n'},
\end{equation}
where 
${\hat \Omega^{\Xi,n(\Theta)}_{\omega, {\bm k}_\perp}} = 
\hat a^{{\rm M},n(\Theta)}_{\omega,{\bm k}_\perp}, 
\hat b^{{\rm M},n(\Theta)}_{\omega,{\bm k}_\perp},
\hat c^{{\rm K},n(\Theta)}_{\omega,{\bm k}_\perp}$ or  
$\hat d^{{\rm K},n(\Theta)}_{\omega,{\bm k}_\perp}$,
with all other anti-commutators between operators in same mode expansion, i.e.\ either in the first or second line of (\ref{ExpandF}), vanishing.
This requirement leads to the following normalization conditions:
\begin{eqnarray}
    (\psi^{\Xi,n(\Theta)}_{\omega, {\bm k}_\perp},\psi^{\Xi,n'(\Theta)}_{\omega', {\bm k'}_\perp})_{\rm D}
    =\int_{\Sigma} d\Sigma_\mu
  \overline{\psi^{\Xi,n(\Theta)}_{\omega, {\bm k}_\perp}}\gamma^\mu_{\rm F(P)}
  \psi^{\Xi,n'(\Theta)}_{\omega', {\bm k}'_\perp}=\delta(\omega-\omega')\delta({\bm k}_\perp-{\bm k}_\perp')\delta_{ n n'}
  .
\label{Fnormal}
\end{eqnarray}
By substituting Eq.~(\ref{anzatzF}) into Dirac equation, we obtain 
\begin{align}
&{\rm F~region~:~}i{\partial\over \partial\eta_{\rm F}}
f^{\Xi,n({\rm F})}_{\omega, {\bm k}_\perp}(\eta_{\rm F}) 
=\biggl[ \gamma^0 me^{a\eta_{\rm F}} -i{a\over2} -\alpha_3\omega+e^{a\eta_{\rm F}}(\alpha_1 k_1+\alpha_2 k_2) \biggr]
f^{\Xi,n({\rm F})}_{\omega, {\bm k}_\perp}(\eta_{\rm F})  ,
\label{DinF}\\
&{\rm P~region~:~}i{\partial\over \partial\eta_{\rm P}}
f^{\Xi,n({\rm P})}_{\omega, {\bm k}_\perp}(\eta_{\rm P}) 
=\biggl[ \gamma^0 me^{-a\eta_{\rm P}}+ i{a\over2} -\alpha_3\omega+e^{-a\eta_{\rm P}}(\alpha_1 k_1+\alpha_2 k_2) \biggr]
f^{\Xi,n({\rm P})}_{\omega, {\bm k}_\perp}(\eta_{\rm P}).
\label{DinP}
\end{align}
We express the solution in the Dirac representation in the following form:
\begin{eqnarray}
f^{\Xi,n(\Theta)}_{\omega, {\bm k}_\perp}(\eta_\Theta)=
\left.
\begin{pmatrix}
\chi^{\Xi,n(\Theta)}_{1,\omega, {\bm k}_\perp}(\eta_\Theta)\\
\chi^{\Xi,n(\Theta)}_{2,\omega, {\bm k}_\perp}(\eta_\Theta)
\end{pmatrix}
\right ..
\label{definition-of-f-in-F}
\end{eqnarray}
Then the Dirac equation gives the following two simultaneous equations:
\begin{align}
&{\rm F~region~:~}
\left\{
\begin{array}{lc}
\displaystyle
i{\partial\over\partial\eta_{\rm F}}\chi^{\Xi,n({\rm F})}_{1,\omega, {\bm k}_\perp}=me^{a\eta_{\rm F}} \chi^{\Xi,n({\rm F})}_{1,\omega, {\bm k}_\perp}-i{a\over 2}\chi^{\Xi,n({\rm F})}_{1,\omega, {\bm k}_\perp}-\omega \sigma^3\chi^{\Xi,n({\rm F})}_{2,\omega, {\bm k}_\perp}+e^{a\eta_{\rm F}} (k_1 \sigma^1+k_2 \sigma^2)\chi^{\Xi,n({\rm F})}_{2,\omega, {\bm k}_\perp} &\\
\displaystyle 
i{\partial\over\partial\eta_{\rm F}}\chi^{\Xi,n({\rm F})}_{2,\omega, {\bm k}_\perp}
=-me^{a\eta_{\rm F}} \chi^{\Xi,n({\rm F})}_{2,\omega, {\bm k}_\perp}-i{a\over 2}\chi^{\Xi,n({\rm F})}_{2,\omega, {\bm k}_\perp}-\omega \sigma^3\chi^{\Xi,n({\rm F})}_{1,\omega, {\bm k}_\perp}+e^{a\eta_{\rm F}} (k_1 \sigma^1+k_2 \sigma^2)\chi^{\Xi,n({\rm F})}_{1,\omega, {\bm k}_\perp}&
\end{array}
\right., \nonumber
\\
&{\rm P~region~:~}
\left\{
\begin{array}{lc}
\displaystyle 
i{\partial\over\partial\eta_{\rm P}}\chi^{\Xi,n({\rm P})}_{1,\omega, {\bm k}_\perp}=me^{-a\eta_{\rm P}} \chi^{\Xi,n({\rm P})}_{1,\omega, {\bm k}_\perp}+i{a\over 2}\chi^{\Xi,n({\rm P})}_{1,\omega, {\bm k}_\perp}-\omega \sigma^3\chi^{\Xi,n({\rm P})}_{2,\omega, {\bm k}_\perp}+e^{-a\eta_{\rm P}} (k_1 \sigma^1+k_2 \sigma^2)\chi^{\Xi,n({\rm P})}_{2,\omega, {\bm k}_\perp}&\\
\displaystyle 
i{\partial\over\partial\eta_{\rm P}}\chi^{\Xi,n({\rm P})}_{2,\omega, {\bm k}_\perp}=-me^{-a\eta_{\rm P}} \chi^{\Xi,n({\rm P})}_{2,\omega, {\bm k}_\perp}+i{a\over 2}\chi^{\Xi,n({\rm P})}_{2,\omega, {\bm k}_\perp}-\omega \sigma^3\chi^{\Xi,n({\rm P})}_{1,\omega, {\bm k}_\perp}+e^{-a\eta_{\rm P}} (k_1 \sigma^1+k_2 \sigma^2)\chi^{\Xi,n({\rm P})}_{1,\omega, {\bm k}_\perp}&
\end{array}
\right.. \nonumber
\end{align}
By combining these simultaneous equations, we obtain the following second order equations:
\begin{align}
&{\rm F~region~:~}{1\over a^2}{\partial^2\over\partial\eta_{\rm F}^2}
\chi^{\Xi,n({\rm F})}_{1\{2\},\omega, {\bm k}_\perp}=-\biggl[{\kappa^2\over a^2}e^{2a\eta_{\rm F}}-{1\over4}+{\omega^2\over a^2}\biggr]\chi^{\Xi,n({\rm F})}_{1\{2\},\omega, {\bm k}_\perp}-{i\omega\over a} \sigma^3\chi^{\Xi,n({\rm F})}_{2\{1\},\omega, {\bm k}_\perp},
\label{2EqF}\\
&{\rm P~region~:~}{1\over a^2}{\partial^2\over\partial\eta_{\rm P}^2}
\chi^{\Xi,n({\rm P})}_{1\{2\},\omega, {\bm k}_\perp}
=-\biggl[{\kappa^2\over a^2}e^{-2a\eta_{\rm P}}-{1\over4}+{\omega^2\over a^2}\biggr]
\chi^{\Xi,n({\rm P})}_{1\{2\},\omega, {\bm k}_\perp}
+{i\omega\over a} \sigma^3\chi^{\Xi,n({\rm P})}_{2\{1\},\omega, {\bm k}_\perp}.
\end{align}
In order to obtain two independent equations from simultaneous equations (\ref{2EqF}), we define:
\begin{eqnarray}
\phi^{\Xi,n(\Theta)}_{\pm,\omega,{\bm k}_\perp}(\eta_\Theta)= 
\chi^{\Xi,n(\Theta)}_{1,\omega, {\bm k}_\perp}
\mp 
\chi^{\Xi,n(\Theta)}_{2,\omega, {\bm k}_\perp}=
\begin{pmatrix}
\varrho^{\Xi,n(\Theta)}_{\pm,\omega,{\bm k}_\perp} (\eta_\Theta) \\
\varsigma^{\Xi,n(\Theta)}_{\pm,\omega,{\bm k}_\perp} (\eta_\Theta)
\end{pmatrix}.
\end{eqnarray}
The signature $\pm$ influences the auxiliary variable in the Bessel function, which construct the solutions, as is shown later.
By using this equation in Eq.~(\ref{definition-of-f-in-F}) we find
\begin{eqnarray}
f^{\Xi,n(\Theta)}_{\omega,{\bm k}_\perp}={1\over2}
\begin{pmatrix}
\varrho^{\Xi,n(\Theta)}_{+,\omega,{\bm k}_\perp}+\varrho^{\Xi,n(\Theta)}_{-,\omega,{\bm k}_\perp} \\
\varsigma^{\Xi,n(\Theta)}_{+,\omega,{\bm k}_\perp}+\varsigma^{\Xi,n(\Theta)}_{-,\omega,{\bm k}_\perp} \\
-\varrho^{\Xi,n(\Theta)}_{+,\omega,{\bm k}_\perp}+\varrho^{\Xi,n(\Theta)}_{-,\omega,{\bm k}_\perp}\\
-\varsigma^{\Xi,n(\Theta)}_{+,\omega,{\bm k}_\perp}+\varsigma^{\Xi,n(\Theta)}_{-,\omega,{\bm k}_\perp} 
\end{pmatrix}.
\end{eqnarray}
Equation~(\ref{2EqF}) written in terms of $\phi^{\Xi,n(\Theta)}_{\pm,\omega,{\bm k}_\perp}(\eta_\Theta)$ is
\begin{align}
&{\rm F~region~:~}{1\over a^2}{\partial^2\over\partial\eta^2_{\rm F}}
\phi^{\Xi,n({\rm F})}_{\pm,\omega,{\bm k}_\perp}
=-\biggl[{\kappa^2\over a^2}e^{2a\eta_{\rm F}}-{1\over4}+{\omega^2 \over a^2}\biggr]
\phi^{\Xi,n({\rm F})}_{\pm,\omega,{\bm k}_\perp}
\pm{i\omega \over a} \sigma^3
\phi^{\Xi,n({\rm F})}_{\pm,\omega,{\bm k}_\perp}, 
\\
&{\rm P~region~:~}{1\over a^2}{\partial^2\over\partial\eta^2_{\rm P}}
\phi^{\Xi,n({\rm P})}_{\pm,\omega,{\bm k}_\perp}
=-\biggl[{\kappa^2\over a^2}e^{-2a\eta_{\rm P}}-{1\over4}+{\omega^2 \over a^2}\biggr]
\phi^{\Xi,n({\rm P})}_{\pm,\omega,{\bm k}_\perp}
\mp{i\omega \over a} \sigma^3
\phi^{\Xi,n({\rm P})}_{\pm,\omega,{\bm k}_\perp}.
\end{align}
Then, we obtain the differential equations,
\begin{align}
&{\rm F~region~:~}\left\{
\begin{array}{lc}
\displaystyle{1\over a^2}{\partial^2\over\partial\eta_{\rm F}^2}
\varrho^{\Xi,n({\rm F})}_{\pm,\omega,{\bm k}_\perp} (\eta_{\rm F})
=\biggl[-{\kappa^2\over a^2}e^{2a\eta_{\rm F}}+\biggl({i\omega \over a}\pm{1\over2}\biggr)^2\biggr]
\varrho^{\Xi,n({\rm F})}_{\pm,\omega,{\bm k}_\perp} (\eta_{\rm F}),&\\
\displaystyle{1\over a^2}{\partial^2\over\partial\eta_{\rm F}^2}
\varsigma^{\Xi,n({\rm F})}_{\pm,\omega,{\bm k}_\perp} (\eta_{\rm F})
=\biggl[-{\kappa\over a^2}e^{2a\eta_{\rm F}}+\biggl({i\omega \over a}\mp{1\over2}\biggr)^2\biggr]
\varsigma^{\Xi,n({\rm F})}_{\pm,\omega,{\bm k}_\perp} (\eta_{\rm F}),&
\end{array}
\right.  
\\
&{\rm P~region~:~}\left\{
\begin{array}{lc}
\displaystyle {1\over a^2}{\partial^2\over\partial\eta_{\rm P}^2}
\varrho^{\Xi,n({\rm P})}_{\pm,\omega,{\bm k}_\perp} (\eta_{\rm P})
=\biggl[-{\kappa^2\over a^2}e^{-2a\eta_{\rm P}}+\biggl({i\omega \over a}\mp{1\over2}\biggr)^2\biggr]
\varrho^{\Xi,n({\rm P})}_{\pm,\omega,{\bm k}_\perp} (\eta_{\rm P}), &\\
\displaystyle {1\over a^2}{\partial^2\over\partial\eta_{\rm P}^2}
\varsigma^{\Xi,n({\rm P})}_{\pm,\omega,{\bm k}_\perp} (\eta_{\rm P})
=\biggl[-{\kappa^2\over a^2}e^{-2a\eta_{\rm P}}+\biggl({i\omega \over a}\pm{1\over2}\biggr)^2\biggr]
\varsigma^{\Xi,n({\rm P})}_{\pm,\omega,{\bm k}_\perp} (\eta_{\rm P}),&
\end{array}
\right.  
\end{align}
which can be solved in terms of the Bessel and related functions.
The solutions of these equations can be written in terms of the Bessel functions as 
\begin{eqnarray}
&&{\rm F~region~:~}\left\{
\begin{array}{lc}
\begin{split}
&\displaystyle  
\varrho^{{\rm K},n({\rm F})}_{+,\omega,{\bm k}_\perp}
=c^{n({\rm F})}_5 J_{\varpi[-i\omega/a-1/2]} (q_{\rm F})~,~~
\varrho^{{\rm K},n({\rm F})}_{-,\omega,{\bm k}_\perp}
=c^{n({\rm F})}_6 J_{\varpi[-i\omega/a+1/2]} (q_{\rm F}),\\
&\displaystyle  \varsigma^{{\rm K},n({\rm F})}_{+,\omega,{\bm k}_\perp}=c^{n({\rm F})}_7 J_{\varpi[-i\omega/a-1/2]} (q_{\rm F})~,~~
\varsigma^{{\rm K},n({\rm F})}_{-,\omega,{\bm k}_\perp}=c^{n({\rm F})}_8 J_{\varpi[-i\omega/ a+1/2]} (q_{\rm F}),
\end{split}
\end{array}
\right. \\
&&{\rm P~region~:~}\left\{
\begin{array}{lc}
\begin{split}
&\displaystyle  
\varrho^{{\rm K},n({\rm P})}_{+,\omega,{\bm k}_\perp}=c^{n({\rm P})}_5 J_{\varpi[i\omega/a+1/2]} (q_{\rm P})~,~~
\varrho^{{\rm K},n({\rm P})}_{-,\omega,{\bm k}_\perp}=c^{n({\rm P})}_6 J_{\varpi[i\omega/a-1/2]} (q_{\rm P}),\\
&\displaystyle  
\varsigma^{{\rm K},n({\rm P})}_{+,\omega,{\bm k}_\perp}=c^{n({\rm P})}_7 J_{\varpi[i\omega/a+1/2]} (q_{\rm P})~,~~
\varsigma^{{\rm K},n({\rm P})}_{-,\omega,{\bm k}_\perp}=c^{n({\rm P})}_8 J_{\varpi[i\omega/a-1/2]} (q_{\rm P}),
\end{split}
\end{array}
\right. 
\label{eq:J-c-coeff}
\end{eqnarray}
or the Hankel function of the second/first kind as
\begin{eqnarray}
&&{\rm F~region~:~}\left\{
\begin{array}{lc}
\begin{split}
&\displaystyle  
\varrho^{{\rm M},n({\rm F})}_{+,\omega,{\bm k}_\perp}
=c^{n({\rm F})}_{5'} H^{(2)}_{\varpi[i\omega/a+1/2]} (q_{\rm F})~,~~
\varrho^{{\rm M},n({\rm F})}_{-,\omega,{\bm k}_\perp}
=c^{n({\rm F})}_{6'} H^{(2)}_{\varpi[i\omega/a-1/2]} (q_{\rm F}),\\
&\displaystyle
\varsigma^{{\rm M},n({\rm F})}_{+,\omega,{\bm k}_\perp}
=c^{n({\rm F})}_{7'} H^{(2)}_{\varpi[i\omega/a+1/2]} (q_{\rm F})~,~~
\varsigma^{{\rm M},n({\rm F})}_{-,\omega,{\bm k}_\perp}
=c^{n({\rm F})}_{8'} H^{(2)}_{\varpi[i\omega/a-1/2]} (q_{\rm F}),
\end{split}
\end{array}
\right.  \\
&&{\rm P~region~:~}\left\{
\begin{array}{lc}
\begin{split}
&\displaystyle  
\varrho^{{\rm M},n({\rm P})}_{+,\omega,{\bm k}_\perp}
=c^{n({\rm P})}_{5'} H^{(1)}_{\varpi[-i\omega/a-1/2]} (q_{\rm P})~,~~
\varrho^{{\rm M},n({\rm P})}_{-,\omega,{\bm k}_\perp}
=c^{n({\rm P})}_{6'} H^{(1)}_{\varpi[-i\omega/a+1/2]} (q_{\rm P}),\\
&\displaystyle  
\varsigma^{{\rm M},n({\rm P})}_{+,\omega,{\bm k}_\perp}
=c^{n({\rm P})}_{7'} H^{(1)}_{\varpi[-i\omega/a-1/2]} (q_{\rm P})~,~~
\varsigma^{{\rm M},n({\rm P})}_{-,\omega,{\bm k}_\perp}
=c^{n({\rm P})}_{8'} H^{(1)}_{\varpi[-i\omega/a+1/2]} (q_{\rm P}),
\end{split}
\end{array}
\right. 
\label{eq:H-c-coeff}
\end{eqnarray}
where
\begin{eqnarray}
q_{\rm F} = \frac{\kappa}{a}e^{a\eta_{\rm F}}~,~~
q_{\rm P} = \frac{\kappa}{a}e^{-a\eta_{\rm P}}.
\label{qFqP}
\end{eqnarray}
Here, we have defined $\varpi=\textrm{sgn}(\omega) = {\omega\over |\omega|}$.  
Each of these two sets of solutions form the ``positive frequency'' subspace of solutions. (See, e.g., Ref.~\cite{QFTCST}
for an explanation of the ``positive frequency'' subspace of solutions.) 
Again, the set of coefficients are not arbitrary, and there are only two
linearly independent spinors. 
We use the following formula to derive the relations between the coefficients in Eq.~(\ref{eq:J-c-coeff}) and (\ref{eq:H-c-coeff}) from the Dirac equation (\ref{DinF}):
\begin{equation}
x\frac{d\ }{dx}Z_\nu(x) = \pm x Z_{\nu\mp 1}(x) \mp \nu Z_\nu(x),
\end{equation}
where $Z_\nu(x) = J_\nu(x)$ , $H^{(1)}_\nu(x)$ or $H^{(2)}_\nu(x)$.

Similarly to the case of the R(L) region, we express the solutions in the following form:
\begin{eqnarray}
f^{\Xi,n(\Theta)}_{\omega, {\bm k}_\perp}(\eta_\Theta)
= A^{\Xi(\Theta)}_{\omega,{\bm k}_\perp}
S^{\Xi,n(\Theta)}_{\omega, {\bm k}_\perp}(\eta_\Theta).
\end{eqnarray}
We note that $ A^{\Xi(\Theta)}_{\omega,{\bm k}_\perp}$ is a real and positive constant which is determined by orthonormality condition (\ref{Fnormal}), 
and $S^{\Xi,n(\Theta)}_{\omega, {\bm k}_\perp}(\eta_\Theta)$ is the spinor which specify the spin state. 
With Bessel function $J_\nu(x)$, we can describe linearly independent and orthogonal spinors can be chosen as follows:
\begin{eqnarray}
&&{\rm F~region~:~}
S^{{\rm K},n({\rm F})}_{\omega, {\bm k}_\perp}(\eta_{\rm F})
=s_{{\bm k}_\perp}^{(+,n)}J_{\varpi[-i\omega/a-1/2]}(q_{\rm F})  -\varpi s_{{\bm k}_\perp}^{(-,n)}J_{\varpi[-i\omega/a+1/2]}(q_{\rm F}),
\label{FS}
 \\
&&{\rm P~region~:~}
S^{{\rm K},n({\rm P})}_{\omega, {\bm k}_\perp}(\eta_{\rm P})
=s_{{\bm k}_\perp}^{(+,n)}J_{\varpi[i\omega/a-1/2]}(q_{\rm P})
+\varpi s_{{\bm k}_\perp}^{(-,n)}J_{\varpi[i\omega/a+1/2]}(q_{\rm P}) .
\label{PS}
\end{eqnarray}
We note that $s_{{\bm k}_\perp}^{(\pm,{\rm u})}$ and  $s_{{\bm k}_\perp}^{(\pm,{\rm d})}$ are already defined by Eq.~(\ref{sdef}) in the previous subsection \ref{Dirac Field in R(L)-Region}.
The normalization constant  is 
\begin{eqnarray}
A^{{\rm K} ({\rm F})}_{\omega,{\bm k}_\perp}=A^{{\rm K} ({\rm P})}_{\omega,{\bm k}_\perp}
={1\over 2\pi} \sqrt{\kappa\over 8a\cosh{\pi\omega/ a}}.
\end{eqnarray} 
We used the formula $J_\nu(z) J_{-\nu+1}(z)+J_{\nu-1}(z) J_{-\nu}(z)=2\sin{(\nu\pi)}/(\pi z) $ 
to determine the normalization constant.
These spinors may be called the Kasner mode in the F(P)-region since it is derived as the solution of Kasner regions, and they construct an orthonormal basis. We can say that these are Rindler modes since we verify that these spinors are virtually connected to Rindler modes in the R and L regions later.
We note that $\omega$ is the energy in the R(L) region satisfying $\omega>0$, while $\omega$ is a momentum-like variable 
in the F(P) region, which can take all real values.
A linearly independent and orthogonal set of two solutions in the F(P) region, 
given in terms of the Hankel function of the second kind can be described as
\begin{eqnarray}
&&{\rm F~region~:~}
S^{{\rm M},n({\rm F})}_{\omega,{\bm k}_\perp}(\eta_{\rm F})
=-i s_{{\bm k}_\perp}^{(+,n)}H^{(2)}_{\varpi [ i\omega/a+1/2 ] }(q_{\rm F}) -i \varpi s_{{\bm k}_\perp}^{(-,n)}H^{(2)}_{ \varpi [ i\omega/a-1/2 ]}(q_{\rm F}), 
\label{FHs}\\
&&{\rm P~region~:~}
S^{{\rm M},n({\rm P})}_{\omega,{\bm k}_\perp}(\eta_{\rm P})
=i s_{{\bm k}_\perp}^{(+,n)}H^{(1)}_{ \varpi [ -i\omega/a+1/2 ]}(q_{\rm P}) -i\varpi s_{{\bm k}_\perp}^{(-,n)}H^{(1)}_{\varpi [-i\omega/a-1/2 ] }(q_{\rm P}),
\label{PHs}
\end{eqnarray}
where the normalization constant is
\begin{eqnarray}
A^{\rm M(F)}_{\omega,{\bm k}_\perp}=A^{\rm M(P)}_{\omega,{\bm k}_\perp}={1\over 8\pi} e^{\pi|\omega|\over2a}\sqrt{\kappa\over a},
\label{FHN}
\end{eqnarray} 
which is determined from the normalization condition and using the mathematical formula 
$H_\nu^{(1)} (z) H_{\nu+1}^{(2)} (z) -H_{\nu+1}^{(1)} (z) H_\nu^{(2)} (z) =4i/(\pi z) $.
Mode functions which construct the normal base with the Hankel functions are called the Minkowski modes (one can find that spinors like Eq.~(\ref{FHs}) and (\ref{PHs}) are certainly the Minkowski mode by the discussion in Sec. \ref{AppendixE}). 
Therefore, annihilation operators associated with the mode functions written with the Hankel functions give the Minkowski vacuum state. We may call them Minkowski mode in the F(P) region.
Mode expansions by the orthonormal base written with Hankel functions and Bessel function $J_\nu(x)$ describe the same Dirac field.
After using the formula $J_{-\nu} (z)={1\over2}~[e^{\nu\pi i}H^{(1)}_\nu (z)+e^{-\nu\pi i}H^{(2)}_\nu (z)]$
,$H^{(1)}_{-\nu}(z)=e^{\nu\pi i}H^{(1)}_\nu(z)$ ,
$H^{(2)}_{-\nu}(z)=e^{-\nu\pi i}H^{(2)}_\nu(z)$
and defining 
\begin{eqnarray}
B_\omega=\sqrt{1\over 2\cosh{\pi\omega/ a}},
\label{defB}
\end{eqnarray}
one obtain the relations of basis of solutions in terms of the Bessel functions and Hankel functions as follows,
\begin{eqnarray}
S^{{\rm K},n(\Theta)}_{\omega, {\bm k}_\perp}(\eta_\Theta)
={\varpi\over 2}e^{{\pi|\omega|\over a}} {S^{{\rm M},n(\Theta)}_{\omega,{\bm k}_\perp}}(\eta_\Theta)
-{1\over2} ({S^{{\rm M},\overline{n}(\Theta)}_{-\omega,-{\bm k}_\perp}})^{\rm C}(\eta_\Theta),
\label{Srelation}
\end{eqnarray}
i.e.
\begin{eqnarray}
\psi^{{\rm K},n(\Theta)}_{\omega,\bm{k}_\perp}=B_\omega \bigl[ \varpi e^{\pi|\omega|\over2a} 
\psi^{{\rm M},n(\Theta)}_{\omega,\bm{k}_\perp}-e^{-{\pi|\omega|\over2a}} 
(\psi^{{\rm M},\overline{n}(\Theta)}_{-\omega,-\bm{k}_\perp})^{\rm C}\bigr],
\label{HJrelationF}
\end{eqnarray}
for $-\infty<\omega<\infty$. 
These relations are nothing but the Bogoliubov transformation because the spinor is associated with the annihilation operator in the mode expansion, and the annihilation operator defines the vacuum state.
Thus, we can express the Bogoliubov transformation in terms of the spinors, operators, and quantum states with the brief deformation. After plugging Eq.~(\ref{HJrelationF}) into the second line of Eq.~(\ref{ExpandF}), we find
\begin{eqnarray}
&&\sum_{n={\rm u,d}} \int_{-\infty} ^{\infty} d\omega \int^\infty_{-\infty} d^2{\bm k}_\perp \biggl( 
{\hat a^{{\rm M},n(\Theta)}_{\omega, {\bm k}_\perp} 
\psi^{{\rm M},n(\Theta)}_{\omega, {\bm k}_\perp} }+
{\hat b^{{\rm M},n(\Theta)\dagger}_{\omega, {\bm k}_\perp} 
(\psi^{{\rm M},n(\Theta)}_{\omega, {\bm k}_\perp} )^{\rm C}} \biggr)\nonumber \\
&=&\sum_{ n={\rm u,d}} \int_{-\infty} ^{\infty} d\omega \int^\infty_{-\infty} d^2{\bm k}_\perp
B_\omega \biggl[  \Bigl(
\varpi e^{\pi|\omega|\over2a} 
\hat c^{{\rm K},n(\Theta)}_{\omega, {\bm k}_\perp} 
-e^{-{\pi|\omega|\over2a}} 
\hat d^{{\rm K},\overline{n}(\Theta)\dagger}_{-\omega, -{\bm k}_\perp,} 
\Bigr)
\psi^{{\rm M},n(\Theta)}_{\omega,\bm{k}_\perp}
\nonumber\\
&&~~~~~~~~~~~~~~~~~~~~~~~~~~~~~~~~~~~~~~~~~~~~
+
\Bigl(
\varpi e^{\pi|\omega|\over2a} 
\hat d^{{\rm K},n(\Theta)\dagger}_{\omega, {\bm k}_\perp}
-e^{-{\pi|\omega|\over2a}} {\hat c^{{\rm K},\overline{n}(\Theta)}_{-\omega, -{\bm k}_\perp}}
\Bigr)
(\psi^{{\rm M},n(\Theta)}_{\omega,\bm{k}_\perp})^{\rm C}
\biggr].
\nonumber
\end{eqnarray}
The comparison between coefficients of mode functions in the first line and the second line indicates the following relation of operators:
\begin{eqnarray}
\hat{a}^{{\rm M},n(\Theta)}_{\omega,\bm{k}_\perp}&=&B_\omega \biggl(
\varpi e^{\pi|\omega|\over2a} 
\hat c^{{\rm K},n(\Theta)}_{\omega, {\bm k}_\perp} 
-e^{-{\pi|\omega|\over2a}} \hat d^{{\rm K},\overline{n}(\Theta)\dagger}_{-\omega, -{\bm k}_\perp} 
\biggr),
\label{Ba}\\
\hat{b}^{{\rm M},n(\Theta)\dagger}_{\omega,\bm{k}_\perp}
&=&B_\omega \biggl(
\varpi e^{\pi|\omega|\over2a} 
\hat d^{{\rm K},n(\Theta)\dagger}_{\omega, {\bm k}_\perp} 
-e^{-{\pi|\omega|\over2a}}
\hat c^{{\rm K}, \overline{n}(\Theta)}_{-\omega, -{\bm k}_\perp}
\biggr).
\label{Bb}
\end{eqnarray}
These formulas can be derived by using the orthonormality of spinors defined with Dirac inner product as following.
We note that there is another way to derive the Bogoliubov transformation, which uses the orthonormality of spinors in the mode expansion. This procedure can be found in other literature (see, e.g. \cite{oriti,Alsing.et.al}).
Here, we briefly review the derivation of the Bogoliubov transformation of operator with orthonormality condition (\ref{Fnormal}) and confirm the consistency.
\begin{eqnarray}
  && \hat a^{{\rm M},n(\Theta)}_{\omega,{\bm k}_\perp}=
  (\psi^{{\rm M},n(\Theta)}_{\omega,{\bm k}_\perp},\hat{\Psi})_{\rm D}
 \nonumber\\
  &&~~~~~~~~=
  \int_{-\infty}^{\infty} d\omega'\int_{-\infty}^{\infty} d^2{\bm k}_\perp'
  \sum_{ n'}\biggl[
    \hat c^{{\rm K},n'(\Theta)}_{\omega',{\bm k}_\perp'} 
    \Bigl(\psi^{{\rm M},n(\Theta)}_{\omega,{\bm k}_\perp},\psi^{{\rm K},n'(\Theta)}_{\omega',{\bm k}_\perp'}\Bigr)_{\rm D}
    + \hat d^{{\rm K},n'(\Theta)\dagger}_{\omega',{\bm k}_\perp'} 
    \Bigl(\psi^{{\rm M},n(\Theta)}_{\omega,{\bm k}_\perp},
    (\psi^{{\rm K},n'(\Theta)}_{\omega',{\bm k}_\perp'})^{\rm C}\Bigr)_{\rm D}
    \biggr],
    \label{Bogoa}
\\
&& \hat b^{{\rm M},n(\Theta)\dagger}_{\omega,{\bm k}_\perp}=\bigl(
(\psi^{{\rm M},n(\Theta)}_{\omega,{\bm k}_\perp})^{\rm C},\hat{\Psi}
\bigr)_{\rm D}
\nonumber\\
&&~~~~~~~~=
  \int_{-\infty}^{\infty} d\omega'\int_{-\infty}^{\infty} d^2{\bm k}_\perp'
  \sum_{ n'}\biggl[
    \hat c^{{\rm K},n'(\Theta)}_{\omega',{\bm k}_\perp'} 
    \Bigl( (\psi^{{\rm M},n(\Theta)}_{\omega,{\bm k}_\perp})^{\rm C} ,\psi^{{\rm K},n'(\Theta)}_{\omega',{\bm k}_\perp'}\Bigr)_{\rm D}
    +\hat d^{{\rm K},n'(\Theta)\dagger}_{\omega',{\bm k}_\perp'} 
    \Bigl( (\psi^{{\rm M},n(\Theta)}_{\omega,{\bm k}_\perp} )^{\rm C} ,
    (\psi^{{\rm K},n'(\Theta)}_{\omega',{\bm k}_\perp'})^{\rm C} \Bigr)_{\rm D} \biggr].
\label{Bogob}
\end{eqnarray}
Thus, Bogoliubov coefficients are derived as the following Dirac inner product:
\begin{eqnarray}
  &&\Bigl(\psi^{{\rm M},n(\Theta)}_{\omega,{\bm k}_\perp},\psi^{{\rm K},n'(\Theta)}_{\omega',{\bm k}_\perp'} \Bigr)_{\rm D}=\varpi B_\omega e^{\pi|\omega|\over2a}
    \delta(\omega-\omega') \delta({\bm k}_\perp-{\bm k}_\perp') \delta_{ n, n'} \label{C1}\\
  &&\Bigl(\psi^{{\rm M},n(\Theta)}_{\omega,{\bm k}_\perp},(\psi^{{\rm K},n'(\Theta)}_{\omega',{\bm k}_\perp'}
  )^{\rm C}\Bigr)_{\rm D}=-B_\omega e^{-{\pi|\omega|\over2a}}
  \delta(\omega+\omega') \delta({\bm k}_\perp+{\bm k}_\perp') \delta_{ n,\overline{n}'} \label{C2}\\
  &&\Bigl((\psi^{{\rm M},n(\Theta)}_{\omega,{\bm k}_\perp})^{\rm  C},\psi^{{{\rm K}},n'(\Theta)}_{\omega',{\bm k}_\perp'}\Bigr)_{\rm D}=-B_\omega e^{-{\pi|\omega|\over2a}}
  \delta(\omega+\omega') \delta({\bm k}_\perp+{\bm k}_\perp') \delta_{ n,\overline{n}'} \label{C3}\\
  &&  \Bigl((\psi^{{\rm M},n(\Theta)}_{\omega,{\bm k}_\perp})^{\rm C},(\psi^{{\rm K,n'}(\Theta)}_{\omega',{\bm k}_\perp'})^{\rm C}\Bigr)_{\rm D}=\varpi B_\omega e^{\pi|\omega|\over2a}
  \delta(\omega-\omega') \delta({\bm k}_\perp-{\bm k}_\perp') \delta_{ n, n'} ,\label{C4}
\end{eqnarray}
where $B_\omega=\bigl[2\cosh{({\pi\omega/a})}\bigr]^{-1/2}$. 
In the calculus of these inner product, the formulae:
$J_{\nu}(z) H^{(1)}_{\nu+1}(z)-J_{\nu+1}(z) H^{(1)}_\nu(z)=-{2i/\pi z}~,~~
J_{\nu}(z) H^{(2)}_{\nu+1}(z)-J_{\nu+1}(z) H^{(2)}_\nu(z)={2i/\pi z}$, may be helpful. By plugging these inner product into the previous relations Eqs. (\ref{Bogoa}), (\ref{Bogob}), we obtain the Bogoliubov transformation which are exactly the same form as Eqs.~(\ref{Ba}) and (\ref{Bb}). 

\section{Analytic Continuation of the Spinors}
\label{Analytic continuation of spinor}
In this section, we demonstrate the procedure of analytic continuation of the spinor solutions. Namely, we clarify how the solutions in the four regions of Minkowski spacetime are related to one another.
The procedure is almost the same as that for a scalar field, as demonstrated in Ref.~\cite{HIUY}.
An additional process is necessary to the spinor field as demonstrated in Appendix
\ref{Lorentz transformation of spinors in Rindler spacetime}.

\subsection{Positive frequency modes for the Minkowski vacuum in the Kasner regions}
\label{AppendixE}

We first demonstrate that the solutions for the Dirac equation in the F region in terms of the Hankel function
of the second kind are indeed the positive-frequency modes for the Minkowski vacuum state.  
We start from the following positive-frequency solutions in the standard coordinate system in the Dirac representation 
with momentum
${\bm k}$:
\begin{equation}
\psi_{\bm k}^{{\rm M},n} = u_{\bm k}^{(n)} e^{-ik_0t + i{\bm k}\cdot {\bm x}},\ \ n={\rm u,d},
\end{equation}
where
\begin{equation}
u^{({\rm u})}_{\bm k} = \frac{1}{\sqrt{2(k_0-k_3)}} \begin{pmatrix} m+k_0 -  k_3\\
-(k_1+ ik_2) \\ m- k_0 + k_3
 \\ k_1+ik_2\end{pmatrix},\ \ \ 
u^{({\rm d})}_{\bm k} = \frac{1}{\sqrt{2(k_0- k_3)}} \begin{pmatrix} k_1-ik_2 \\  m+k_0 - k_3 \\ k_1 - ik_2 \\ 
-m+k_0 - k_3\end{pmatrix}.
\end{equation}
Here, $k_0$, $k_1$, $k_2$, $k_3$ are the \textit{contravariant} components, i.e., they are the $4$-momentum
components with the upper indices.  These solutions satify the following normalization conditions:
\begin{eqnarray}
(\psi_{\bm k}^{{\rm M},n},\psi_{{\bm k}'}^{{\rm M},n'})_{\rm D} & = & 
\int d^3{\bm x} \psi_{\bm k}^{{\rm M},n\dagger}\psi_{{\bm k}'}^{{\rm M},n'} \nonumber \\
& =&  2k_0 (2\pi)^3 \delta^{nn'}\delta({\bm k}-{\bm k}').
\end{eqnarray}
Next we define the global-defined solutions
\begin{eqnarray}
\Psi_{\omega,{\bm k}_\perp}^{{\rm M},n}  
& = & \frac{i^{1/2}}{\sqrt{2a}(2\pi)^{2}} \int_{-\infty}^\infty
\frac{dk_3}{k_0} \left( \frac{k_0 + k_3}{k_0 - k_3}\right)^{-i \omega/2a}
u_{\bm k}^{(n)} e^{-ik_0 t + i {\bm k}\cdot {\bm x}}. \label{eq:global-solution}
\end{eqnarray}
Here the parameter $\omega$ takes any real value.
We change the integration variable from $k_3$ to the rapidity,
\begin{equation}
\theta = \frac{1}{2}\log \frac{k_0 +k_3}{k_0 - k_3}.
\end{equation}
Then these solutions can be written as
\begin{equation}
\Psi^{{\rm M},n}_{\omega,{\bm k}_\perp}(t,z,{\bm x}_\perp) 
= \left[ v_{{\bm k}_\perp}^{(+,n)} F_{\omega,{\bm k}_\perp}^{(+)}(t,z)
+ v_{{\bm k}_\perp}^{(-,n)} F_{\omega,{\bm k}_\perp}^{(-)}(t,z)\right] e^{i{\bm k}_\perp\cdot {\bm x}_\perp},
\label{eq:PsiMn}
\end{equation}
where
\begin{eqnarray}
&&v_{{\bm k}_\perp}^{(+,{\rm u})}  =  \frac{1}{8\pi^2\sqrt{a\kappa}} 
\Bigl(\kappa,0,-\kappa,0\Bigr)^T=  \frac{1}{8\pi^2}\sqrt{\kappa\over a}s^{(+,{\rm u})}_{\bm k_\perp},\\
&&v_{{\bm k}_\perp}^{(-,{\rm u})} =  \frac{1}{8\pi^2\sqrt{a\kappa}} 
\biggl(m,-(k_1+ik_2),m,k_1+ik_2\biggr)^T=-\frac{i}{8\pi^2}\sqrt{\kappa\over a}s^{(-,{\rm u})}_{\bm k_\perp} ,\\
&&v_{{\bm k}_\perp}^{(+,{\rm d})} 
= \frac{1}{8\pi^2\sqrt{a\kappa}} \Bigl(0,\kappa,0,\kappa\Bigr)^T=\frac{i}{8\pi^2}\sqrt{\kappa\over a}s^{(+,{\rm d})}_{\bm k_\perp} ,\\
&&v_{{\bm k}_\perp}^{(-,{\rm d})} = \frac{1}{8\pi^2 \sqrt{a\kappa}} \Bigl(k_1-ik_2,m,k_1-ik_2,-m\Bigr)^T=\frac{1}{8\pi^2}\sqrt{\kappa\over a}s^{(-,{\rm d})}_{\bm k_\perp},
\end{eqnarray}
and
\begin{equation}
F^{(\pm)}_{\omega,{\bm k}_\perp}(t,z) = i^{1/2}\int_{-\infty}^\infty d\theta\, e^{-(i\omega/a \pm 1/2)\theta}
\exp \left( - i(\kappa \cosh \theta)t + i (\kappa \sinh \theta)z\right).  \label{eq:F-definition}
\end{equation}
The solutions $\Psi_{\omega,{\bm k}_\perp}^{{\rm M},n}$ are normalized as
\begin{equation}
(\Psi_{\omega,{\bm k}_\perp}^{{\rm M},n},\Psi_{\omega',{\bm k}'_{\perp}}^{{\rm M},n'})_{\rm D}
= \delta^{nn'}\delta(\omega-\omega')\delta({\bm k}_\perp - {\bm k}'_\perp).
\end{equation}
We note that
the charge conjugation of the spinor $\Psi$ is defined by
$\Psi^{\rm C} = i\gamma^2\Psi^*$. 

It is useful for later purposes to discuss some properties of the constant spinors $v_{{\bm k}_\perp}^{(\pm,n)}$.  
By defining $\overline{\rm u}={\rm d}$ and $\overline{\rm d}={\rm u}$, we find
\begin{eqnarray}
v_{-{\bm k}_\perp}^{(+,n){\rm C}} = v_{{\bm k}_\perp}^{(+,\overline{n})},\label{eq:conj-v1}\\
v_{-{\bm k}_\perp}^{(-,n){\rm C}} = - v_{{\bm k}_\perp}^{(-,\overline{n})}. \label{eq:conj-v2}
\end{eqnarray}
We also note that $\alpha_3 v_{{\bm k}_\perp}^{(\pm,n)} = \mp v_{{\bm k}_\perp}^{(\pm,n)}$. Hence
\begin{equation}
\exp(b\alpha_3) v_{{\bm k}_\perp}^{(\pm,n)} = \exp(\mp b) v_{{\bm k}_\perp}^{(\pm,n)}, \label{eq:alpha3}
\end{equation}
for any number $b$.  It is also useful to note that the multiplication by $\exp(b\alpha_3)$ and the charge conjugation
commute because $[\alpha_3,\gamma_{\rm M}^2]=0$ and because $\alpha_3$ is a real matrix.

It is now straightforward to express the solutions $\Psi_{\omega,{\bm k}_\perp}^{{\rm M},n}$ in each four regions.
We start with the F and P regions.  We first find the functions $F^{(\pm)}_{\omega,{\bm k}_\perp}$ defined by 
Eq.~(\ref{eq:F-definition}) by substituting 
$(t,z)=(a^{-1}e^{a\eta_{\rm F}}\cosh a\zeta_{\rm F},a^{-1}e^{a\eta_{\rm F}}\sinh a\zeta_{\rm F})$ in the
F region and $(t,z)=(a^{-1}e^{-a\eta_{\rm P}} \cosh a\zeta_{\rm P},a^{-1}e^{-a\eta_{\rm P}}\sinh a\eta_{\rm P})$
in the P region and by using the following formulas (8.421 of Ref.~\cite{Gradshteyn}):
\begin{eqnarray}
\int_{-\infty}^\infty e^{-ix\cosh \theta - \nu\theta}d\theta
& = & - \pi i e^{-\nu\pi i/2}H_\nu^{(2)}(x),\\
\int_{-\infty}^\infty e^{ix\cosh \theta - \nu\theta}d\theta
& = & \pi i e^{-\nu\pi i/2}H_{-\nu}^{(1)}(x), \label{eq:H1}
\end{eqnarray}
for $|{\textrm Re}\,\nu| < 1$ and $x$ real.  We also 
used the fact that $e^{-\nu\pi i/2}H^{(1)}_{\nu}(x)$ is even in $\nu$ in Eq.~(\ref{eq:H1}).  
Let the function $F^{(\pm)}_{\omega,{\bm k}_\perp}$ in the F and P regions
be denoted by $F^{({\rm F},\pm)}_{\omega,{\bm k}_\perp}$ and $F^{({\rm P},\pm)}_{\omega,{\bm k}_\perp}$ respectively.
Then, we find
\begin{eqnarray}
F^{({\rm F},+)}_{\omega,{\bm k}_\perp}
& = & -i\pi e^{\pi\omega/2a} e^{-i\omega \zeta_{\rm F}} e^{-a\zeta_{\rm F}/2} H^{(2)}_{i\omega/a+1/2}
(q_{\rm F}), \label{eq:FFplus}\\
F^{({\rm F},-)}_{\omega,{\bm k}_\perp}
& = & \pi e^{\pi\omega/2a} e^{-i\omega\zeta_{\rm F}} e^{a\zeta_{\rm F}/2}H^{(2)}_{i\omega/a -1/2}(q_{\rm F}),\\
F^{({\rm P},+)}_{\omega,{\bm k}_\perp}
& = & i\pi e^{\pi\omega/2a} e^{i\omega \zeta_{\rm P}} e^{a\zeta_{\rm P}/2} H^{(1)}_{-i\omega/a-1/2}
(q_{\rm P}), \\
F^{({\rm P},-)}_{\omega,{\bm k}_\perp}
& = & -\pi e^{\pi\omega/2a} e^{i\omega\zeta_{\rm P}} 
e^{-a\zeta_{\rm P}/2}H^{(1)}_{-i\omega/a+1/2}(q_{\rm P}). 
\label{eq:FPminus}
\end{eqnarray}

The $\gamma$-matrices for 
the spinor components obtained by 
substituting Eqs.~(\ref{eq:FFplus})-(\ref{eq:FPminus}) into Eq.~(\ref{eq:PsiMn})  correspond
to the cartesian coordinates.  As discussed in Appendix~\ref{Lorentz transformation of spinors in Rindler spacetime},
these spinors expressed in the coordinate systems used for the F and P regions, 
denoted by $\Psi_{\omega,{\bm k}_\perp}^{{\rm M},n({\rm F})}$ and $\Psi_{\omega,{\bm k}_\perp}^{{\rm M},n({\rm P})}$,
respectively, are
\begin{eqnarray}
\Psi_{\omega,{\bm k}_\perp}^{{\rm M},n({\rm F})} & = & e^{-a\zeta_{\rm F}\alpha_3/2}
\Psi_{\omega,{\bm k}_\perp}^{{\rm M},n} \nonumber \\
& = &  \left[ v_{{\bm k}_\perp}^{(+,n)} e^{a\zeta_{\rm F}/2}F_{\omega,{\bm k}_\perp}^{({\rm F},+)}(t,z)
+ v_{{\bm k}_\perp}^{(-,n)} e^{-a\zeta_{\rm F}/2}
F_{\omega,{\bm k}_\perp}^{({\rm F}-)}\right] e^{i{\bm k}_\perp\cdot {\bm x}_\perp},\\
\Psi_{\omega,{\bm k}_\perp}^{{\rm M},n({\rm P})} & = & e^{a\zeta_{\rm F}\alpha_3/2}
\Psi_{\omega,{\bm k}_\perp}^{{\rm M},n} \nonumber \\
& = &  \left[ v_{{\bm k}_\perp}^{(+,n)} e^{-a\zeta_{\rm F}/2}F_{\omega,{\bm k}_\perp}^{({\rm P},+)}
+ v_{{\bm k}_\perp}^{(-,n)} e^{a\zeta_{\rm F}/2}
F_{\omega,{\bm k}_\perp}^{({\rm P},-)}\right] e^{i{\bm k}_\perp\cdot {\bm x}_\perp},
\end{eqnarray}
where we have used Eq.~(\ref{eq:alpha3}).
Here, by using the formulae, $H_\nu^{(2)}(x)=e^{\nu\pi i/2}H_{-\nu}^{(2)}(x)$ and $H_\nu^{(1)}(x)=e^{-\nu\pi i/2}H_{-\nu}^{(1)}(x)$, we obtain 
\begin{eqnarray}
\Psi_{\omega,{\bm k}_\perp}^{{\rm M},n({\rm F})}
& = & \pi e^{\pi \omega/2a} e^{-i\omega\zeta_{\rm F}}
\left[-i v_{{\bm k}_\perp}^{(+,n)}H^{(2)}_{i\omega/a+1/2}(q_{\rm F})
+ v_{{\bm k}_\perp}^{(-,n)}H^{(2)}_{i\omega/a - 1/2}(q_{\rm F})\right]
e^{i{\bm k}_\perp\cdot {\bm x}_\perp} \nonumber\\
& =  &  \pi e^{-\pi \omega/2a} e^{-i\omega\zeta_{\rm F}}
\left[ v_{{\bm k}_\perp}^{(+,n)}H^{(2)}_{-(i\omega/a+1/2)}(q_{\rm F})
- i v_{{\bm k}_\perp}^{(-,n)}H^{(2)}_{-(i\omega/a - 1/2)}(q_{\rm F})\right]
e^{i{\bm k}_\perp\cdot {\bm x}_\perp},\\
\Psi_{\omega,{\bm k}_\perp}^{{\rm M},n({\rm P})}
&=& -\pi e^{\pi \omega/2a} e^{i\omega\zeta_{\rm P}}
\left[ -iv_{{\bm k}_\perp}^{(+,n)}H^{(1)}_{-i\omega/a-1/2}(q_{\rm P})
+v_{{\bm k}_\perp}^{(-,n)}H^{(1)}_{-i\omega/a + 1/2}(q_{\rm P})\right]
e^{i{\bm k}_\perp\cdot {\bm x}_\perp} \nonumber\\
 & = &  -\pi e^{-\pi \omega/2a} e^{i\omega\zeta_{\rm P}}
\left[ v_{{\bm k}_\perp}^{(+,n)}H^{(1)}_{i\omega/a+1/2}(q_{\rm P})
-i v_{{\bm k}_\perp}^{(-,n)}H^{(1)}_{i\omega/a - 1/2}(q_{\rm P})\right]
e^{i{\bm k}_\perp\cdot {\bm x}_\perp}.
\end{eqnarray}
Hence the charge conjugation is
\begin{eqnarray}
\Psi^{{\rm M},n({\rm F}){\rm C}}_{-\omega,-{\bm k}_\perp}
&=&-\pi e^{-\pi\omega/2a} e^{-i\omega\zeta_{\rm F}}
\left[
-i v^{(+,\overline{n})}_{\bm k} H^{(1)}_{i\omega/a+1/2}(q_{\rm F})
+v^{(-,\overline{n})}_{\bm k} H^{(1)}_{i\omega/a-1/2}(q_{\rm F})
\right] e^{i{\bm k}_\perp \cdot {\bm x}_\perp} \nonumber\\
&=&\pi e^{\pi\omega/2a} e^{-i\omega\zeta_{\rm F}}
\left[
v^{(+,\overline{n})}_{\bm k} H^{(1)}_{-(i\omega/a+1/2)}(q_{\rm F})
-i v^{(-,\overline{n})}_{\bm k} H^{(1)}_{-(i\omega/a-1/2)}(q_{\rm F})
\right] e^{i{\bm k}_\perp \cdot {\bm x}_\perp} ,\\
\Psi^{{\rm M},n({\rm P}){\rm C}}_{-\omega,-{\bm k}_\perp}
&=& \pi e^{-\pi\omega/2a} e^{i\omega\zeta_{\rm P}}
\left[
-iv^{(+,\overline{n})}_{\bm k} H^{(2)}_{-i\omega/a-1/2}(q_{\rm P})
+v^{(-,\overline{n})}_{\bm k} H^{(2)}_{-i\omega/a+1/2}(q_{\rm P})
\right] e^{i{\bm k}_\perp \cdot {\bm x}_\perp} \nonumber\\
&=& -\pi e^{\pi\omega/2a} e^{i\omega\zeta_{\rm P}}
\left[
v^{(+,\overline{n})}_{\bm k} H^{(2)}_{i\omega/a+1/2}(q_{\rm P})
-i v^{(-,\overline{n})}_{\bm k} H^{(2)}_{i\omega/a-1/2}(q_{\rm P})
\right] e^{i{\bm k}_\perp \cdot {\bm x}_\perp} .
\end{eqnarray}
We note that these results give similar forms as the spinors in Sec. \ref{Dirac Field in F-Kasner Region}, i.e. they are related by
\begin{eqnarray}
&\Psi_{\omega,{\bm k}_\perp}^{{\rm M},{\rm u}({\rm F})} = \psi^{\rm M,u(F)}_{\omega,{\bm k}_\perp} (\omega >0),~~ 
&\Psi_{\omega,{\bm k}_\perp}^{{\rm M},{\rm d}({\rm F})} = i \psi^{\rm M,d(F)}_{\omega,{\bm k}_\perp} (\omega >0),\\ 
&\Psi_{\omega,{\bm k}_\perp}^{{\rm M},{\rm u}({\rm F})}= i \psi^{\rm M,u(F)}_{\omega,{\bm k}_\perp} (\omega < 0),~~
&\Psi_{\omega,{\bm k}_\perp}^{{\rm M},{\rm d}({\rm F})}=-\psi^{\rm M,d(F)}_{\omega,{\bm k}_\perp} (\omega < 0),
\\
&\Psi_{\omega,{\bm k}_\perp}^{{\rm M},{\rm u}({\rm P})} = \psi^{\rm M,u(P)}_{-\omega,{\bm k}_\perp} (\omega >0), ~~
&\Psi_{\omega,{\bm k}_\perp}^{{\rm M},{\rm d}({\rm P})} = - \psi^{\rm M,d(P)}_{-\omega,{\bm k}_\perp} (\omega >0),\\ 
&\Psi_{\omega,{\bm k}_\perp}^{{\rm M},{\rm u}({\rm P})} = i \psi^{\rm M,u(P)}_{-\omega,{\bm k}_\perp} (\omega <0), ~~
&\Psi_{\omega,{\bm k}_\perp}^{{\rm M},{\rm d}({\rm P})} =- \psi^{\rm M,d(P)}_{-\omega,{\bm k}_\perp} (\omega <0).
\end{eqnarray}

\subsection{Analytic continuation}
Subsequently, we consider the analytic continuation
of the spinors in the F region and the P region
into those in the R region and the L region.
The coordinate variables of the four regions are related by the analytic continuation in Table~I. 
This table is applicable only for the positive-frequency solutions in Minkowski spacetime.
As is shown in the case of the scalar field, 
one needs to take into account the difference in the analytic
properties between the positive frequency and negative frequency modes at the boundaries of the four regions where they are singular~\cite{HIUY}. 
In addition, the components of the spinors analytically continued from one region to another do not satisfy the Dirac equation in the latter region.
The reason is that the components of the spinors depend on the tetrads, which are chosen differently in the four regions.
We have to consider the continuation of the spinor solutions taking the direction of the tetrads into account, which is explained in Appendix A.
This gives an additional complication to make the relation between the spinor solutions in the four regions compared with the case of a scalar field.

Let us express 
the spinor solutions $\Psi^{{\rm M},n}_{\omega,{\bm k}_\perp}$ in the R and L regions.  We substitute
$(t,z) = (a^{-1}e^{a\xi_{\rm R}}\sinh a\tau_{\rm R}, a^{-1}e^{a\xi_{\rm R}} \\ \cosh a \tau_{\rm R}$) in the
R region and $(t,z) = (a^{-1}e^{a\xi_{\rm L}}\sin a\tau_{\rm L}, -a^{-1}e^{a\xi_{\rm L}}\cosh a\tau_{\rm L})$
in the L region into Eq.~(\ref{eq:F-definition})
to find  $F^{(\pm)}_{\omega,{\bm k}_\perp}$ using 8.432.1 in Ref.~\cite{Gradshteyn}: 
\begin{equation}
\int_{-\infty}^\infty e^{-\nu\theta}e^{-x\cosh\theta}\,d\theta 
= 2K_\nu(x).
\end{equation}
Let the functions $F^{(\pm)}_{\omega,{\bm k}_\perp}$ in the R and L regions be denoted by
$F^{({\rm R},\pm)}_{\omega,{\bm k}_\perp}$ and $F^{({\rm L},\pm)}_{\omega,{\bm k}_\perp}$, respectively.
Then
\begin{eqnarray}
F_{\omega,{\bm k}_\perp}^{({\rm R},+)}
& = & 2e^{\pi \omega/2a} e^{-i\omega \tau_{\rm R}}e^{- a\tau_{\rm R}/2}K_{i\omega/a+1/2}(q_{\rm R}),\\
F_{\omega,{\bm k}_\perp}^{({\rm R},-)}
& = & 2ie^{\pi \omega/2a} e^{-i\omega \tau_{\rm R}}e^{a\tau_{\rm R}/2}K_{i\omega/a-1/2}(q_{\rm R}),\\
F^{({\rm L},+)}_{\omega,{\bm k}_\perp} & = & 2i
e^{-\pi\omega/2a} e^{i\omega \tau_{\rm L}} e^{a\tau_{\rm L}/2}K_{i\omega/a+1/2}(q_{\rm L}),\\
F^{({\rm L},-)}_{\omega,{\bm k}_\perp} & = & 2
e^{-\pi\omega/2a} e^{i\omega \tau_{\rm L}} e^{-a\tau_{\rm L}/2}K_{i\omega/a-1/2}(q_{\rm L}).
\end{eqnarray}
Then the spinors $\Psi_{\omega,{\bm k}_\perp}^{{\rm M},n}$ can readily be found 
using Eq.~(\ref{eq:PsiMn}).  We need to make corrections due to different choices of the $\gamma$-matrices, which depends on the tetrad, 
before comparing them to the solutions found in Sec.~II. Thus, we define spinors that can be compared to those found in the R and L regions (see Appendix~\ref{Lorentz transformation of spinors in Rindler spacetime}) as
\begin{eqnarray}
\Psi_{\omega,{\bm k}_\perp}^{{\rm M},n({\rm R})} & = & \exp(- a\tau_{\rm R}\alpha_3/2)
\Psi^{{\rm M},n}_{\omega,{\bm k}_\perp},\\
\Psi_{\omega,{\bm k}_\perp}^{{\rm M},n({\rm L})} & = &\gamma^3\gamma_5 \exp( a\tau_{\rm L}\alpha_3/2)
\Psi^{{\rm M},n}_{\omega,{\bm k}_\perp} \nonumber \\
& =  & \exp( - a\tau_{\rm L}\alpha_3/2)
\gamma^3\gamma_5\Psi^{{\rm M},n}_{\omega,{\bm k}_\perp},  \label{eq:gamma-reversed}
\end{eqnarray}
respectively.  The second expression in Eq.~(\ref{eq:gamma-reversed}) follows because
the matrices $\gamma^3\gamma_5 = \textrm{diag}(1,-1,-1,1)$ and $\alpha_3$ anti-commute.  
 Thus, we obtain
\begin{eqnarray}
\Psi_{\omega,{\bm k}_\perp}^{{\rm M},n({\rm R})} & = &  2e^{\pi \omega/2a} e^{-i\omega\tau_{\rm R}}
\left[ v_{{\bm k}_\perp}^{(+,n)}K_{i\omega/a+1/2}(q_{\rm R})
+iv_{{\bm k}_\perp}^{(-,n)}K_{i\omega/a - 1/2}(q_{\rm R})\right]
e^{i{\bm k}_\perp\cdot {\bm x}_\perp},\\
\Psi_{\omega,{\bm k}_\perp}^{{\rm M},n({\rm L})} &=&
 2 e^{-\pi \omega/2a} e^{i\omega\tau_{\rm L}}
\gamma^3\gamma_5 \left[i v_{{\bm k}_\perp}^{(+,n)}K_{i\omega/a+1/2}(q_{\rm L})
+v_{{\bm k}_\perp}^{(-,n)}K_{i\omega/a - 1/2}(q_{\rm L})\right]
e^{i{\bm k}_\perp\cdot {\bm x}_\perp}.
\end{eqnarray}
We note that the following relation is useful to find relations to the solutions 
in Sec.~II. 
\begin{eqnarray}
&&\gamma^3\gamma_5v_{{\bm k}_\perp}^{(+,{\rm u})}  =  \frac{1}{8\pi^2\sqrt{a\kappa}} 
\Bigl(\kappa,0,\kappa,0\Bigr)^T=  \frac{-i}{8\pi^2}\sqrt{\kappa\over a}\tilde{s}^{(-,{\rm u})}_{\bm k_\perp},\\
&&\gamma^3\gamma_5v_{{\bm k}_\perp}^{(-,{\rm u})} =  \frac{1}{8\pi^2\sqrt{a\kappa}} 
\biggl(m,k_1+ik_2,-m,k_1+ik_2\biggr)^T=\frac{1}{8\pi^2}\sqrt{\kappa\over a}\tilde{s}^{(+,{\rm u})}_{\bm k_\perp} ,\\
&&\gamma^3\gamma_5v_{{\bm k}_\perp}^{(+,{\rm d})} 
= \frac{1}{8\pi^2\sqrt{a\kappa}} \Bigl(0,-\kappa,0,\kappa\Bigr)^T=\frac{i}{8\pi^2}\sqrt{\kappa\over a}\tilde{s}^{(-,{\rm d})}_{\bm k_\perp} ,\\
&&\gamma^3\gamma_5v_{{\bm k}_\perp}^{(-,{\rm d})} = \frac{1}{8\pi^2 \sqrt{a\kappa}} \Bigl(k_1-ik_2,-m,-(k_1-ik_2),-m\Bigr)^T=-\frac{1}{8\pi^2}\sqrt{\kappa\over a}\tilde{s}^{(+,{\rm d})}_{\bm k_\perp}.
\end{eqnarray}   

By using Eqs.~(\ref{eq:conj-v1}) and (\ref{eq:conj-v2}) we find
\begin{eqnarray}
\Psi_{-\omega,{\bm k}_\perp}^{{\rm M},n({\rm R})} & = & e^{-\pi\omega/a} 
(\Psi_{\omega,-{\bm k}_\perp}^{{\rm M},\overline{n}({\rm R}) })^{\rm C},\\
\Psi_{-\omega,{\bm k}_\perp}^{{\rm M},n({\rm L})} & = & - e^{\pi\omega/a}
(\Psi_{\omega,-{\bm k}_\perp}^{{\rm M},\overline{n}({\rm L})})^{\rm C}.
\end{eqnarray}
Thus, we find that globally-defined solutions $\Psi_{\omega,{\bm k}_\perp}^{{\rm M},n}$ given by
Eq.~(\ref{eq:global-solution}) become in the R and L regions
\begin{equation}
\Psi_{\omega,{\bm k}_\perp}^{{\rm M}, n(\Lambda)} \rightarrow 
\begin{cases} \Psi_{\omega,{\bm k}_\perp}^{{\rm M},n({\rm R})} &  (\textrm{R region}) \\
-e^{-\pi\omega/a}(\Psi^{{\rm M},\overline{n}({\rm L})}_{-\omega,-{\bm k}_\perp})^{\rm C}=\Psi^{{\rm M},n({\rm L})}_{\omega,{\bm k}_\perp} & (\textrm{L region}) 
\end{cases},
\end{equation}
\begin{equation}
\Psi_{-\omega,{\bm k}_\perp}^{{\rm M}, n(\Lambda)}  \rightarrow
\begin{cases} e^{-\pi\omega/a}(\Psi_{\omega,-{\bm k}_\perp}^{{\rm M},\overline{n}({\rm R})})^{\rm C}=\Psi^{{\rm M},n({\rm R})}_{-\omega,{\bm k}_\perp} &  (\textrm{R region}) \\
\Psi^{{\rm M},n({\rm L})}_{-\omega,{\bm k}_\perp} & (\textrm{L region}) \end{cases}.
\end{equation}

We define the global modes, which are defined in all the regions by analytic continuation 
and by making the correction due to the difference of the tetrad, by
\begin{eqnarray}
\Psi^{{\rm I},n(X)}_{|\omega|,{\bm k}_\perp}=B_\omega\Bigl( e^{\pi|\omega|/2a} \Psi^{{\rm M},n(X)}_{|\omega|,{\bm k}_\perp} + e^{-\pi|\omega|/2a} (\Psi^{{\rm M},\overline{n}(X)}_{-|\omega|,-{\bm k}_\perp})^{\rm C} \Bigr),
\label{RM}
\\
\Psi^{{\rm II},n(X)}_{|\omega|,{\bm k}_\perp}=B_\omega \Bigl( e^{\pi|\omega|/2a} \Psi^{{\rm M},n(X)}_{-|\omega|,{\bm k}_\perp} - e^{-\pi|\omega|/2a} (\Psi^{{\rm M},\overline{n}(X)}_{|\omega|,-{\bm k}_\perp})^{\rm C} \Bigr), 
\label{LM}
\end{eqnarray}
with $X=$ F,~P,~(R/L) to denote different expressions in each region.

In the R and L region, we find that the functions
(\ref{RM}) and (\ref{LM}) reduce to
\begin{eqnarray}
\Psi^{{\rm I},n({\rm R/L})}_{|\omega|,{\bm k}_\perp}
=\left\{   \begin{array}{lc}
\displaystyle 
\sqrt{1+e^{-2\pi|\omega|/a}}~
\Psi^{{\rm M},n({\rm R})}_{|\omega|,{\bm k}_\perp}  &{\rm (R~region)}\\    \displaystyle \\
\displaystyle 
0  &{\rm (L~region)}
\end{array}    \right.
\label{AppendixpsiRn}
\end{eqnarray}
and
\begin{eqnarray}
\Psi^{{\rm II},n({\rm R/L})}_{|\omega|,{\bm k}_\perp}
=\left\{   \begin{array}{lc}
\displaystyle 
0 &{\rm (R~region)}\\    \displaystyle \\
\displaystyle 
\sqrt{1+e^{-2\pi|\omega|/a}}~
\Psi^{{\rm M},n({\rm L})}_{-|\omega|,{\bm k}_\perp}  &{\rm (L~region)}
\end{array}    \right.
\label{AppendixpsiLn}
\end{eqnarray}
In the F region, from Eqs.~(\ref{RM}) and (\ref{LM}),
 we find that these modes yield
\begin{eqnarray}
&&\Psi^{{\rm I},n({\rm F})}_{|\omega|,{\bm k}_\perp}=2\pi B_\omega e^{-i|\omega|\zeta_{\rm F}} \biggl[ v^{(+,n)}_{\bm k_\perp} J_{-(i|\omega|/a+1/2)}(q_{\rm F})-iv^{(-,n)}_{\bm k_\perp} J_{-(i|\omega|/a-1/2)}(q_{\rm F}) \biggr] e^{i{\bm k}_\perp\cdot {\bm x}_\perp},
\\
&&\Psi^{{\rm II},n({\rm F})}_{|\omega|,{\bm k}_\perp}=2\pi B_\omega e^{i|\omega|\zeta_{\rm F}} \biggl[-iv^{(+,n)}_{\bm k_\perp} J_{-i|\omega|/a+1/2}(q_{\rm F})+v^{(-,n)}_{\bm k_\perp} J_{-i|\omega|/a-1/2}(q_{\rm F}) \biggr] e^{i{\bm k}_\perp\cdot {\bm x}_\perp}.
\end{eqnarray}
for $\omega>0$ and $\omega<0$, respectively. 
In the P region, Eqs.~(\ref{RM}) and (\ref{LM}) give
\begin{eqnarray}
&&\Psi^{{\rm I},n({\rm P})}_{|\omega|,{\bm k}_\perp}=-2\pi B_\omega e^{-i|\omega|\zeta_{\rm P}} \biggl[-iv^{(+,n)}_{\bm k_\perp} J_{i|\omega|/a-1/2}(q_{\rm P})+v^{(-,n)}_{\bm k_\perp} J_{i|\omega|/a+1/2}(q_{\rm P}) \biggr] e^{i{\bm k}_\perp\cdot {\bm x}_\perp}, 
\label{AppendixPR} \\
&&\Psi^{{\rm II},n({\rm P})}_{|\omega|,{\bm k}_\perp}=-2\pi B_\omega e^{i|\omega|\zeta_{\rm P}} \biggl[ v^{(+,n)}_{\bm k_\perp} J_{i|\omega|/a+1/2}(q_{\rm P})-iv^{(-,n)}_{\bm k_\perp} J_{i|\omega|/a-1/2}(q_{\rm P}) \biggr] e^{i{\bm k}_\perp\cdot {\bm x}_\perp}.
\label{AppendixPL}
\end{eqnarray}
for $\omega<0$ and $\omega>0$, respectively.

\subsection{Summary of the Analytic continuation}
In the previous subsection, we introduced the global modes 
by Eqs.~(\ref{RM}) and (\ref{LM}),  
whose explicit expressions are given by (\ref{AppendixpsiRn})--(\ref{AppendixPL}).
The global modes are the solutions analytically continued 
in the entire region of Minkowski spacetime and corrected the difference due
to the tetrad. We find the relations to the solutions found 
in Sec.~II, which are summarized as follows:
\begin{eqnarray}
&&\psi^{\rm I,u}_{\omega,\bm{k}_\perp}
\equiv\left\{ \begin{array}{l}
\displaystyle 
\Psi^{\rm I,u(F)}_{\omega,{\bm k}_\perp}=\psi^{\rm K,u(F)}_{\omega,\bm{k}_\perp}~~~~~~(\omega>0) ~~{\rm F}\\
\displaystyle 
\Psi^{\rm I,u(R/L)}_{\omega,{\bm k_\perp}}=
\left\{
\begin{array}{lcc}
\psi^{\rm R,u}_{\omega,\bm{k}_\perp} &(\omega>0)&{\rm R}\\
0~~~~~~&(\omega>0)&{\rm L}\\
\end{array}
\right.
\\
\displaystyle 
 \Psi^{\rm I,u(P)}_{\omega,{\bm k}_\perp}=i\psi^{\rm K,u(\rm P)}_{\omega,\bm{k}_\perp}~~~~~(\omega>0)~~{\rm P}
\end{array}
\right. 
~,~~~
\psi^{\rm I,d}_{\omega,\bm{k}_\perp}
\equiv\left\{
\begin{array}{l}
\displaystyle 
\Psi^{\rm I,d(F)}_{\omega,{\bm k}_\perp}=i \psi^{\rm K,d(F)}_{\omega,\bm{k}_\perp}~~~~~~~(\omega>0)~~{\rm F}\\
\displaystyle 
\Psi^{\rm I,d(R/L)}_{\omega,{\bm k_\perp}}=
\left\{
\begin{array}{lcc}
\displaystyle
i\psi^{\rm R,d}_{\omega,\bm{k}_\perp} ~&(\omega>0)&{\rm R}\\
\displaystyle
0&(\omega>0)&{\rm L}\\
\end{array}
\right.
\\
\displaystyle 
\Psi^{\rm I,d(P)}_{\omega,{\bm k}_\perp}=-\psi^{\rm K,d(P)}_{\omega,\bm{k}_\perp}~~~~~~(\omega>0)~~{\rm P}
\end{array}
\right.,
\label{PsiI,IIu}
\end{eqnarray}
and
\begin{eqnarray}
&&\psi^{\rm II,u}_{|\omega|,\bm{k}_\perp}
\equiv\left\{
\begin{array}{l}
\displaystyle 
\Psi^{\rm II,u(F)}_{-\omega,-{\bm k}_\perp}=-i\psi^{\rm K,u(F)}_{\omega,-\bm{k}_\perp} ~~~~(\omega<0)~~{\rm F}\\
\Psi^{\rm II,u(R/L)}_{-\omega,-{\bm k}_\perp}=
\left\{
\begin{array}{lcc}
\displaystyle
0&(\omega<0)&{\rm R}\\
\displaystyle 
\psi^{\rm L,u}_{-\omega,-\bm{k}_\perp} &(\omega<0)&{\rm L}\\
\end{array}
\right.
\\
\displaystyle 
\Psi^{\rm II,u(P)}_{-\omega,-{\bm k}_\perp}=-\psi^{\rm K,u(P)}_{\omega,-\bm{k}_\perp} ~~~~~~(\omega<0)~~{\rm P}
\end{array}
\right.  
~,~~~
\psi^{\rm II,d}_{|\omega|,\bm{k}_\perp}
\equiv\left\{
\begin{array}{l}
\displaystyle 
\Psi^{\rm II,d(F)}_{-\omega,-{\bm k}_\perp}=\psi^{\rm K,d(F)}_{\omega,-\bm{k}_\perp} ~~~~~~~~~~(\omega<0)~~{\rm F}\\
\Psi^{\rm II,d(R/L)}_{-\omega,-{\bm k}_\perp}=
\left\{
\begin{array}{lcc}
\displaystyle 
0&(\omega<0)&{\rm R}\\
\displaystyle 
-\psi^{\rm L,d}_{-\omega,-\bm{k}_\perp} &(\omega<0)&{\rm L}\\
\end{array}
\right.
\\
\displaystyle 
\Psi^{\rm II,d(P)}_{-\omega,-{\bm k}_\perp}=-i \psi^{\rm K,d(P)}_{\omega,-\bm{k}_\perp} ~~~~~~~(\omega<0)~~{\rm P}
\end{array}
\right.
\label{PsiI,IIb}
\end{eqnarray}
or equivalently,
\begin{eqnarray}
&&\psi^{\rm II,u}_{\omega,\bm{k}_\perp}
\equiv\left\{
\begin{array}{l}
\displaystyle 
\Psi^{\rm II,u(F)}_{\omega,-{\bm k}_\perp}=-i\psi^{\rm K,u(F)}_{-\omega,-\bm{k}_\perp} ~~~~(\omega>0)~~{\rm F}\\
\Psi^{\rm II,u(R/L)}_{\omega,-{\bm k}_\perp}=
\left\{
\begin{array}{lcc}
\displaystyle
0&(\omega>0)&{\rm R}\\
\displaystyle 
\psi^{\rm L,u}_{\omega,-\bm{k}_\perp} ~~&(\omega>0)&{\rm L}\\
\end{array}
\right.
\\
\displaystyle 
\Psi^{\rm II,u(P)}_{\omega,-{\bm k}_\perp}=-\psi^{\rm K,u(P)}_{-\omega,-\bm{k}_\perp} ~~~~~~(\omega>0)~~{\rm P}
\end{array}
\right.  
~,~~~
\psi^{\rm II,d}_{\omega,\bm{k}_\perp}
\equiv\left\{
\begin{array}{l}
\displaystyle 
\Psi^{\rm II,d(F)}_{\omega,-{\bm k}_\perp}=\psi^{\rm K,d(F)}_{-\omega,-\bm{k}_\perp} ~~~~~~~~~~(\omega>0)~~{\rm F}\\
\Psi^{\rm II,d(R/L)}_{\omega,-{\bm k}_\perp}=
\left\{
\begin{array}{lcc}
\displaystyle 
0&(\omega>0)&{\rm R}\\
\displaystyle 
-\psi^{\rm L,d}_{\omega,-\bm{k}_\perp} \ ~&(\omega>0)&{\rm L}\\
\end{array}
\right.
\\
\displaystyle 
\Psi^{\rm II,d(P)}_{\omega,-{\bm k}_\perp}=-i \psi^{\rm K,d(P)}_{-\omega,-\bm{k}_\perp} ~~~~~~~(\omega>0)~~{\rm P}
\end{array}
\right.  .
\label{PsiI,II}
\end{eqnarray}
We note again that $\Psi^{\vartheta,n(X)}_{\omega,\bm k_\perp}$
is constructed by the function $\Psi^{{\rm M},n}_{\omega, \bm k_\perp}$ 
in Eq.~(\ref{eq:global-solution}), which is globally defined
by a linear combination of the positive-frequency modes
of the Minkowski coordinates, 
and by applying the local Lorentz transformation from the local Lorentz frame of 
the Minkowski coordinates to that in each local 
coordinates of the regions F, P, L, R, to take the difference of the tetrad into 
account. 
Therefore, $\psi^{\vartheta,n}_{\omega,\bm k_\perp}$ 
represents the solutions analytically continued across different region, whose explicit expressions are summarized in Appendix C.

The Dirac field operator in the F region is written as 
\begin{eqnarray}
{\hat \psi}(x)
&=&\sum_{ n={\rm u,d}} \int_{-\infty} ^{\infty} d\omega \int^\infty_{-\infty} d^2{\bm k}_\perp \biggl( {\hat c^{{\rm K},n({\rm F})}_{\omega, {\bm k}_\perp} \psi^{{\rm K},n({\rm F})}_{\omega, {\bm k}_\perp} (x)}+
{\hat d^{{\rm K},n({\rm F})\dagger}_{\omega, {\bm k}_\perp} (\psi^{{\rm K}, n({\rm F})}_{\omega, {\bm k}_\perp} (x))^{\rm C}} \biggr)\label{FmodeE}\\
\nonumber\\
&=&\sum_{ n={\rm u,d}} \int_{0} ^{\infty} d\omega \int^\infty_{-\infty} d^2{\bm k}_\perp \biggl({\hat c^{{\rm K},n({\rm F})}_{\omega, {\bm k}_\perp} \psi^{{\rm K}, n({\rm F})}_{\omega, {\bm k}_\perp} (x)}+{\hat d^{{\rm K},n({\rm F})\dagger}_{\omega, {\bm k}_\perp} (\psi^{{\rm K},n({\rm F})}_{\omega, {\bm k}_\perp} (x))^{\rm C}} \biggr)
\nonumber\\
&&+\sum_{ n={\rm u,d}} \int_{0} ^{\infty} d\omega \int^\infty_{-\infty} d^2{\bm k}_\perp \biggl({\hat c^{{\rm K},n({\rm F})}_{-\omega, -{\bm k}_\perp} \psi^{{\rm K}, n({\rm F})}_{-\omega,-{\bm k}_\perp} (x)}+
{\hat d^{{\rm K},n({\rm F})\dagger}_{-\omega, -{\bm k}_\perp} (\psi^{{\rm K},n({\rm F})}_{-\omega,-{\bm k}_\perp} (x))^{\rm C}} \biggr).
\label{fieldpsiF}
\end{eqnarray}
Here, $x$ denotes the coordinate $(t,{\bm x})$ in the Minkowski spacetime. From the behaviour of the solution near the Rindler horizon in the F region, we refer $\psi_{\omega,\bm k_\perp}^{{\rm K},n({\rm F})}(x)$ with $\omega>0$ and $\psi_{\omega,\bm k_\perp}^{{\rm K}, n({\rm F})}(x)$ with $\omega<0$ 
the left-moving wave modes and the right-moving wave modes, respectively.
Similarly, we refer $\psi_{\omega,\bm k_\perp}^{{\rm K},n({\rm P})}(x)$ with 
$\omega>0$ and $\psi_{\omega,\bm k_\perp}^{{\rm K},n({\rm P})}(x)$ with $\omega<0$ the right-moving wave modes and the left-moving wave modes from the behaviour near the Rindler horizon 
in the P region.
The results of the analytic continuation show that the 
left-moving wave modes in the F region and the 
right-moving wave modes in the P region 
are analytically continued to the Rindler modes in the R region and to zero in the L region, 
which are represented by $\psi^{{\rm I},n}_{\omega,\bm k_\perp}$. 
Furthermore, the right-moving wave modes in the F region 
and the left-moving wave modes in the P-region are
analytically continued to the Rindler modes in the L region and to 
zero in the R region, which are represented by $\psi^{{\rm II},n}_{\omega,\bm k_\perp}$.
Thus, the expression of the Dirac field operator (\ref{fieldpsiF}) 
can be extended to the 
entire region of Minkowski spacetime as 
\begin{align}
{\hat \psi}(x)
=\sum_{\vartheta={\rm I,II}}  \sum_{ n={\rm u,d}} \int_{0} ^{\infty} d\omega \int^\infty_{-\infty} d^2{\bm k}_\perp \biggl( 
{\hat c^{\vartheta, n}_{\omega, {\bm k}_\perp} \psi^{\vartheta, n}_{\omega, {\bm k}_\perp} (x)}+{\hat d^{\vartheta, n\dagger}_{\omega, {\bm k}_\perp} (\psi^{\vartheta, n}_{\omega, {\bm k}_\perp} (x))^{\rm C}} \biggr),
\label{I,IImode}
\end{align}
where the mode functions characterized by the index ``I'' and ``II'' are defined by Eqs.~(\ref{PsiI,IIu})
and (\ref{PsiI,II}), and the creation and the annihilation operator satisfy the anti-commutation relations
\begin{equation}
\{ {\hat c^{\vartheta,n}_{\omega, {\bm k}_\perp}} , {\hat c^{\vartheta',n'\dagger}_{\omega', {\bm k}_\perp'}} \}
=\{ {\hat d^{\vartheta,n}_{\omega, {\bm k}_\perp}} , {\hat d^{\vartheta',n'\dagger}_{\omega', {\bm k}_\perp'}} \}
=\delta(\omega-\omega')\delta({\bm k}_\perp-{\bm k}_\perp')\delta_{ n n'}\delta_{\vartheta\vartheta'},
\end{equation}
with all other anti-commutators vanishing.
Comparing the expressions of the Dirac field operator,  Eqs. (\ref{fieldpsiF})
and (\ref{I,IImode}) with the use of 
 the expressions (\ref{PsiI,IIu}) and (\ref{PsiI,II}),
we have the following relations in the F region
\begin{eqnarray}
\left.
\begin{array}{l}
\displaystyle 
\hat{c}^{\rm K,u(F)}_{\omega,\bm{k}_\perp}=
\hat{c}^{\rm I,u}_{\omega,\bm{k}_\perp}\\
\hat{d}^{\rm K,u(F)\dagger}_{\omega,\bm{k}_\perp}=
\hat{d}^{\rm I,u\dagger}_{\omega,\bm{k}_\perp}\\
\end{array}
\right.,
\left.
\begin{array}{l}
\displaystyle 
\hat{c}^{\rm K,d(F)}_{\omega,\bm{k}_\perp}=
i\hat{c}^{\rm I,d}_{\omega,\bm{k}_\perp}\\
\hat{d}^{\rm K,d(F)\dagger}_{\omega,\bm{k}_\perp}=
-i\hat{d}^{\rm I,d\dagger}_{\omega,\bm{k}_\perp}\\
\end{array}
\right.,
\left.
\begin{array}{l}
\displaystyle 
\hat{c}^{\rm K,u(F)}_{-\omega,-\bm{k}_\perp}=
-i\hat{c}^{\rm II,u}_{\omega,\bm{k}_\perp}\\
\hat{d}^{\rm K,u(F)\dagger}_{-\omega,-\bm{k}_\perp}=
i\hat{d}^{\rm II,u\dagger}_{\omega,\bm{k}_\perp}\\
\end{array}
\right.,
\left.
\begin{array}{l}
\displaystyle 
\hat{c}^{\rm K,d(F)}_{-\omega,-\bm{k}_\perp}=
\hat{c}^{\rm II,d}_{\omega,\bm{k}_\perp}\\
\hat{d}^{\rm K,d(F)\dagger}_{-\omega,-\bm{k}_\perp}=
\hat{d}^{\rm II,d\dagger}_{\omega,\bm{k}_\perp}\\
\end{array}
\right.,
\label{I,II-K}
\end{eqnarray}
and in the P region
\begin{eqnarray}
\left.
\begin{array}{l}
\displaystyle 
\hat{c}^{\rm K,u(P)}_{\omega,\bm{k}_\perp}=
i\hat{c}^{\rm I,u}_{\omega,\bm{k}_\perp}\\
\hat{d}^{\rm K,u(P)\dagger}_{\omega,\bm{k}_\perp}=
-i\hat{d}^{\rm I,u\dagger}_{\omega,\bm{k}_\perp}\\
\end{array}
\right.,
\left.
\begin{array}{l}
\displaystyle 
\hat{c}^{\rm K,d(P)}_{\omega,\bm{k}_\perp}=
-\hat{c}^{\rm I,d}_{\omega,\bm{k}_\perp}\\
\hat{d}^{\rm K,d(P)\dagger}_{\omega,\bm{k}_\perp}=
-\hat{d}^{\rm I,d\dagger}_{\omega,\bm{k}_\perp}\\
\end{array}
\right.,
\left.
\begin{array}{l}
\displaystyle 
\hat{c}^{\rm K,u(P)}_{-\omega,-\bm{k}_\perp}=
-\hat{c}^{\rm II,u}_{\omega,\bm{k}_\perp}\\
\hat{d}^{\rm K,u(P)\dagger}_{-\omega,-\bm{k}_\perp}=
-\hat{d}^{\rm II,u\dagger}_{\omega,\bm{k}_\perp}\\
\end{array}
\right.,
\left.
\begin{array}{l}
\displaystyle 
\hat{c}^{\rm K,d(P)}_{-\omega,-\bm{k}_\perp}=
-i\hat{c}^{\rm II,d}_{\omega,\bm{k}_\perp}\\
\hat{d}^{\rm K,d(P)\dagger}_{-\omega,-\bm{k}_\perp}=
i\hat{d}^{\rm II,d\dagger}_{\omega,\bm{k}_\perp}\\
\end{array}
\right..
\label{I,II-K2}
\end{eqnarray}

\begin{figure}[h]
  \includegraphics[width=90mm]{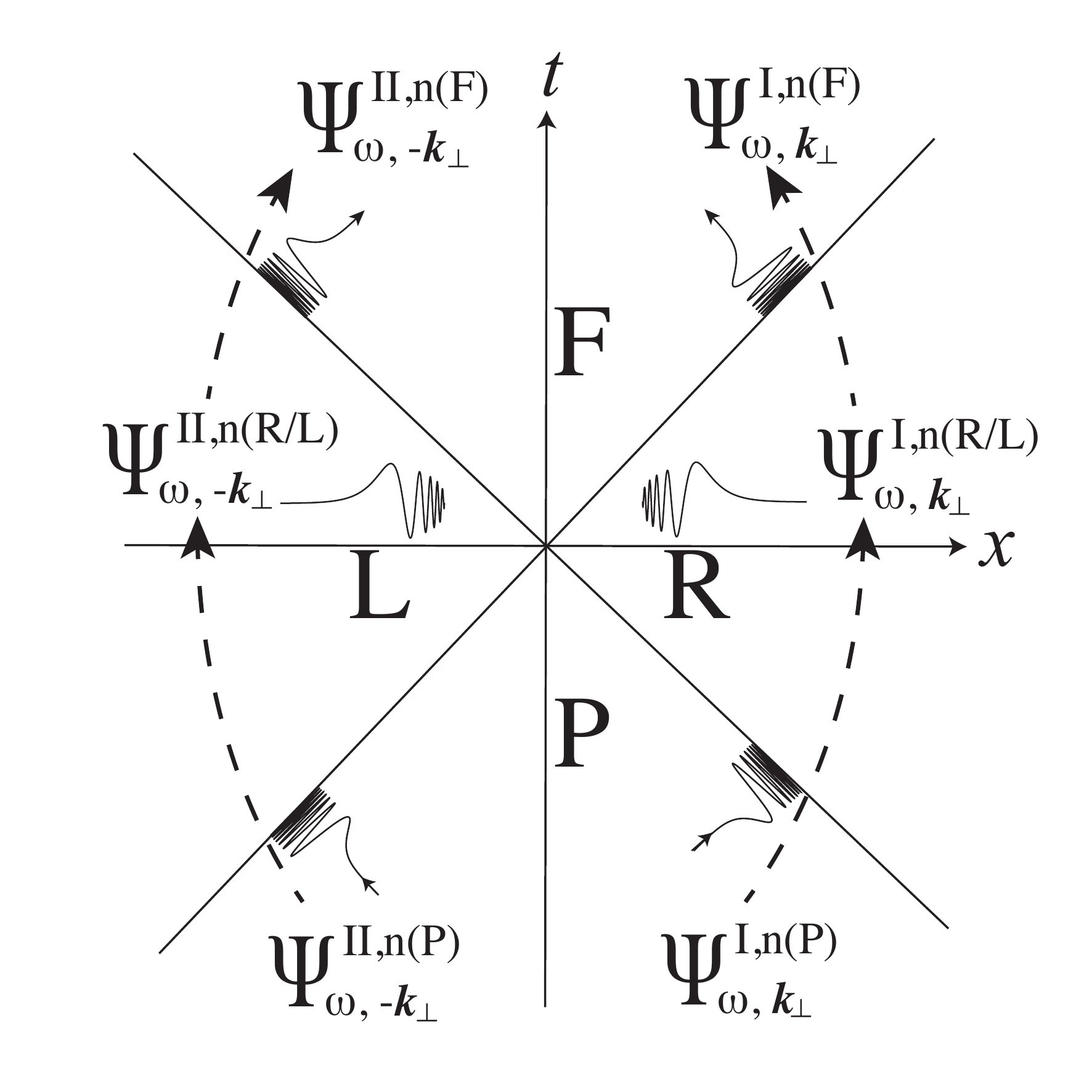}
\caption{
Relations of the mode functions in each region. 
This figure shows the equivalence of $\Psi^{\rm I,n(F)}_{\omega,{\bm k}_\perp}$(left-\\ moving wave modes in the F region),  
$\Psi^{\rm I,n(R/L)}_{\omega,{\bm k}_\perp}$(right Rindler modes), and $\Psi^{\rm I,n(P)}_{\omega,{\bm k}_\perp}$(right-moving wave \\modes in the P region). 
Similarly, $\Psi^{\rm II,n(F)}_{\omega,-{\bm k}_\perp}$(right-moving wave modes in the F region), $\Psi^{\rm II,n(R/L)}_{\omega,-{\bm k}_\perp}$(left \\ Rindler modes), 
 $\Psi^{\rm II,n(P)}_{\omega,-{\bm k}_\perp}$(left-moving wave modes in the P region) are equivalent. ~~~~~~~~~~~~~~~~~~~~~~~
 \label{modes2}}
\end{figure}
\section{Expression of the Minkowski vacuum state and the Unruh effect}
\label{Discussion}
The Minkowski vacuum state of a scalar field is described by an entangled state (see e.g.~\cite{HIUY}, 
cf.~\cite{SYK} for the gravitational wave). 
We focus on the Minkowski vacuum sate of a Dirac field
using the analytic continuation-property of the general 
4-dimensional Dirac spinor developed in the previous section.
We first focus on the vacuum structure in Kasner regions, which can be extended to the entire region of the Minkowski spacetime. We also discuss the Unruh effect of the Dirac field after discussing the description of the Minkowski vacuum.

In order to derive the Minkowski vacuum state for the Dirac field as an entangled state, we start with adopting the anzatz for the Minkowski vacuum state 
\begin{eqnarray}
|0^{\rm M}\rangle =\prod_{\omega\gtrless0}\prod_{~\bm {\bm k}_\perp}\prod_{ n}|0^{{\rm M}, n}_{\omega,{\bm k}_\perp}\rangle
\label{Vacuum State0}
\end{eqnarray}
in Kasner regions with
\begin{eqnarray}
|0^{{\rm M}, n}_{\omega,{\bm k}_\perp}\rangle =\sum_{\ell=0}^{1}C_\ell|\ell^{{\rm K}, n(\Theta)}_{\omega,{\bm k}_\perp}\rangle^c|\ell^{{\rm K},\overline{n}(\Theta)}_{-\omega,-{\bm k}_\perp}\rangle^d, 
\label{Vacuum State}
\end{eqnarray}
where $|\ell^{{\rm K}, n(\Theta)}_{\omega,{\bm k}_\perp}\rangle^c$ and 
$|\ell^{{\rm K},\overline{n}(\Theta)}_{-\omega,-{\bm k}_\perp}\rangle^d$ denote the ground state ($\ell=0$) or the one particle state ($\ell=1$) for a particle of the index "$c$" with the momentum $(\omega,\bm k_\perp)$
and for an anti-particle of the index "$d$" with the momentum
$(-\omega,-\bm k_\perp)$. Namely, we may define
the ground state by
\begin{eqnarray}
\hat c^{{\rm K},n(\Theta)}_{\omega,{\bm k}_\perp}|0^{{\rm K},n(\Theta)}_{\omega,{\bm k}_\perp}\rangle^c=0~,~~
\hat d^{{\rm K},n(\Theta)}_{\omega,{\bm k}_\perp}|0^{{\rm K},n(\Theta)}_{\omega,{\bm k}_\perp}\rangle^d=0
\end{eqnarray}
and we write the one particle excited particle state ($\ell=1$) for particle and anti-particle 
\begin{eqnarray}
|1^{{\rm K},n(\Theta)}_{\omega,{\bm k}_\perp}\rangle^c=\hat c^{{\rm K}, n(\Theta)\dagger}_{\omega,{\bm k}_\perp}|0^{{\rm K},n(\Theta)}_{\omega,{\bm k}_\perp}\rangle^c~,~~
~~|1^{{\rm K},n(\Theta)}_{\omega,{\bm k}_\perp}\rangle^d=\hat d^{{\rm K}, n(\Theta)\dagger}_{\omega,{\bm k}_\perp}|0^{{\rm K},n(\Theta)}_{\omega,{\bm k}_\perp}\rangle^d. 
\end{eqnarray}
The ansatz Eq.~(\ref{Vacuum State0}) comes from the anti-commutation relation to obtain the maximally entangled state (see Appendix \ref{The ansatz of the Minkowski vacuum state for fermion and anti-fermion}). 

Now we find the relation of the states, which are related by the Bogoliubov transformation
(\ref{Ba}) and (\ref{Bb}). 
To find the coefficient $C_\ell$ in the ansatz Eq.~(\ref{Vacuum State}), we substitute the ansatz 
and the Bogoliubov transformation into the relation $\hat{a}^{{\rm M}, n(\Theta)}_{\omega,{\bm k}_\perp}|0^{{\rm M}, n}_{\omega,{\bm k}_\perp}\rangle=0$, which leads to
\begin{eqnarray}
\hat{a}^{{\rm M},n(\Theta)}_{\omega,{\bm k}_\perp}|0^{{\rm M},n}_{\omega,{\bm k}_\perp}\rangle
&=&B_\omega 
\Bigl(\varpi e^{\pi|\omega|/2a}
\hat c^{{\rm K},n(\Theta)}_{\omega, {\bm k}_\perp} 
-e^{-{\pi|\omega|/2a}}
\hat d^{{\rm K},\overline{n}(\Theta)\dagger}_{-\omega, -{\bm k}_\perp} 
\Bigr)
\sum_{\ell=0}^{1}C_\ell|\ell^{{\rm K},n(\Theta)}_{\omega,{\bm k}_\perp}\rangle^c|\ell^{{\rm K},\overline{n}(\Theta)}_{-\omega,-{\bm k}_\perp}\rangle^d  \nonumber\\
&=&B_\omega
\Bigl(
\varpi e^{\pi|\omega|/2a} 
C_1 - e^{-{\pi|\omega|/2a}} 
C_0 \Bigr)
|0^{{\rm K},n(\Theta)}_{\omega,{\bm k}_\perp} \rangle^c|1^{{\rm K},\overline{n}(\Theta)}_{-\omega,-{\bm k}_\perp}\rangle^d=0.
\end{eqnarray}
The result gives the following relation for the coefficients
\begin{eqnarray}
C_1=\varpi e^{-{\pi|\omega|/ a}} C_0.
\label{Crelation}
\end{eqnarray}
Therefore we find that the Minkowski vacuum state is written in the form (\ref{Vacuum State0}) with
\begin{align}
|0^{{\rm M},n}_{\omega,{\bm k}_\perp}\rangle
={1\over\sqrt{e^{-{2\pi|\omega|/a}}+1}}
 \Bigl(
|0^{{\rm K},n(\Theta)}_{\omega,{\bm {\bm k}_\perp}}\rangle^c |0^{{\rm K},\overline{n}(\Theta)}_{-\omega,-{\bm {\bm k}_\perp}}\rangle^d
+\varpi e^{-{\pi|\omega|/ a}}
|1^{{\rm K},n(\Theta)}_{\omega,{\bm {\bm k}_\perp}}\rangle^c |1^{{\rm K},\overline{n}(\Theta)}_{-\omega,-{\bm {\bm k}_\perp}}\rangle^d 
\Bigl),
\label{vstate}
\end{align}
where the constant $C_0$ was determined by the normalization condition, $\langle 0^{{\rm M}, n}_{\omega,{\bm k}_\perp}|0^{{\rm M}, n}_{\omega,{\bm k}_\perp}\rangle=1$, as $
|C_0|^2={[e^{-{2\pi|\omega|/ a}}+1]^{-1}}$.
Furthermore, with the fact
\begin{eqnarray}
\hat{b}^{{\rm M},n(\Theta)}_{\omega,{\bm k}_\perp}|0^{{\rm M},\bar n}_{-\omega,-{\bm k}_\perp}\rangle
&=&{B_\omega \over \sqrt{e^{-2\pi|\omega|/a}+1}} 
\Bigl(
\varpi e^{\pi|\omega|/2a} 
\hat d^{{\rm K},n(\Theta)}_{\omega, {\bm k}_\perp} 
-e^{-{\pi|\omega|/2a}} 
{\hat c}^{{\rm K},n(\Theta)\dagger}_{\omega, {\bm k}_\perp} \Bigr)
\nonumber\\
&&
\Bigl(
|0^{{\rm K},\overline{n}(\Theta)}_{-\omega,-{\bm {\bm k}_\perp}}\rangle^c |0^{{\rm K},n(\Theta)}_{\omega,{\bm k}_\perp}\rangle^d
-\varpi
e^{-{\pi|\omega|/ a}}
|1^{{\rm K},\overline{n}(\Theta)}_{-\omega,-{\bm {\bm k}_\perp}}\rangle^c |1^{{\rm K},n(\Theta)}_{\omega,{\bm {\bm k}_\perp}}\rangle^d 
\Bigl)
  \nonumber\\
&=&0,
\end{eqnarray}
where we used the anti-commutation relation between $\hat{d}^{{\rm K},n(\Theta)}_{\omega,\bm{k}_\perp}$ and $\hat{c}^{{\rm K},\overline{n}(\Theta)}_{-\omega,-\bm{k}_\perp}$ to derive the second equality, we conclude that 
Eq.~(\ref{Vacuum State0}) with (\ref{vstate}) is indeed the Minkowski vacuum state for anti-fermions too.
It is important to note that the expression of the vacuum state is valid in both of F-region and P region. 
Thus the Minkowski vacuum of a Dirac field is described by the entangled state between particle 
labelled by `` $c$ " and anti-particle labelled by `` $d$ " with momenta opposite to each other
in the F region as well as in the P region.

Using the equivalence of the mode functions presented in the last part of the previous section, 
i.e., the equivalence between the left-moving (right-moving) wave mode in the F region and the right (left) Rinder mode and the right-moving (left-moving) wave mode in the P region, we can extend the expression to the modes labelled by the indices I and II.   
Because Eq.~(\ref{Vacuum State0}) is rewritten as
\begin{eqnarray}
|0^{\rm M}\rangle &=&\prod_{\omega\gtrless0}\prod_{~\bm {\bm k}_\perp}\prod_{ n}|0^{{\rm M}, n}_{\omega,{\bm k}_\perp}\rangle
\nonumber\\
&=&\prod_{\omega>0}\prod_{~\bm {\bm k}_\perp}\prod_{ n}
{1\over e^{-{2\pi\omega/a}}+1}
\Bigl(|0^{{\rm K},n(\Theta)}_{\omega,{\bm k}_\perp} \rangle^c |0^{{\rm K},\overline{n}(\Theta)}_{-\omega,-{\bm k}_\perp}\rangle^d
+e^{-{\pi\omega/a}}|1^{{\rm K},n(\Theta)}_{\omega,{\bm k}_\perp} \rangle^c |1^{{\rm K},\overline{n}(\Theta)
}_{-\omega,-{\bm k}_\perp}\rangle^d
\Bigr)\nonumber\\
&&~~~~~~~~~~~~~\otimes \Bigl(|0^{{\rm K},n(\Theta)}_{-\omega,{\bm k}_\perp} \rangle^c 
|0^{{\rm K},\overline{n}(\Theta)}_{\omega,-{\bm k}_\perp}\rangle^d
-e^{-{\pi\omega/ a}}|1^{{\rm K},n(\Theta)}_{-\omega,{\bm k}_\perp} \rangle^c |1^{{\rm K},\overline{n}(\Theta)}_{\omega,-{\bm k}_\perp}\rangle^d \Bigr),
\label{zeroM}
\end{eqnarray}
therefore we can rewrite the Minkowski vacuum state of the Dirac field ~(\ref{Vacuum State0}) 
as
\begin{eqnarray}
|0^{\rm M}\rangle
&=&\prod_{\omega>0}\prod_{~~\bm {\bm k}_\perp}\prod_{ n}
{1\over e^{-{2\pi\omega/a}}+1}
\Bigl(|0^{{\rm I},n}_{\omega,{\bm k}_\perp} \rangle^c |0^{{\rm II},\overline{n}}_{\omega,-{\bm k}_\perp}\rangle^d
+e^{-{\pi\omega/ a}}|1^{{\rm I},n}_{\omega,{\bm k}_\perp} \rangle^c |1^{{\rm II},\overline{n}
}_{\omega,-{\bm k}_\perp}\rangle^d
\Bigr)\nonumber\\
&&~~~~~~~~~~~~~\otimes 
\Bigl(|0^{{\rm II},n}_{\omega,{\bm k}_\perp} \rangle^c |0^{{\rm I},\overline{n}}_{\omega,-{\bm k}_\perp}\rangle^d
-e^{-{\pi\omega/ a}}|1^{{\rm II},n}_{\omega,{\bm k}_\perp} \rangle^c |1^{{\rm I},\overline{n}}_{\omega,-{\bm k}_\perp}\rangle^d \Bigr), 
\label{zeroMM}
\end{eqnarray}
where we defined the ground state by
\begin{eqnarray}
\hat c^{\vartheta,n}_{\omega,\bm k_\perp}|0_{\omega,\bm k_\perp}^{\vartheta,n}\rangle^c=0, ~~
\hat d^{\vartheta,n}_{\omega,\bm k_\perp}|0_{\omega,\bm k_\perp}^{\vartheta,n}\rangle^d=0
\end{eqnarray}
and the one particle excited state by
\begin{eqnarray}
|1_{\omega,\bm k_\perp}^{\vartheta,n}\rangle^c=\hat c^{\vartheta,n\dagger}_{\omega,\bm k_\perp}|0_{\omega,\bm k_\perp}^{\vartheta,n}\rangle^c, 
~~
|1_{\omega,\bm k_\perp}^{\vartheta,n}\rangle^d=\hat d^{\vartheta,n\dagger}_{\omega,\bm k_\perp}|0_{\omega,\bm k_\perp}^{\vartheta,n}\rangle^d.
\end{eqnarray}
In the derivation of Eq.~(\ref{zeroMM}) from Eq.~(\ref{zeroM}), we used the fact that the modes labeled by $(-\omega>0,\bm k_\perp)$ in the F region correspond to the modes
II with $(\omega>0,\bm k_\perp)$, 
which comes from the relations 
(\ref{PsiI,II}), as well as the relations (\ref{I,II-K}).
The modes I and II are defined in the entire Minkowski spacetime, and that the modes labelled by I vanish in the L region, while other modes labelled by II vanish in the R region. This is the important property of the four-dimensional case even for the massless case. This is the contrasting property to the two-dimensional
massless case \cite{Alsing.et.al}. These properties are common to those of a scalar field \cite{HIUY}.

Using the above results, let us discuss the Unruh effect of a Dirac field. 
An observer in the R region is disconnected to the L region, thus we take the partial trace over the density matrix of the Minkowski vacuum with respect to the modes in the L region specified by II
\begin{eqnarray}
  \hat \rho^{\rm I}&=&
  {\rm Tr_{II}} \bigl[ |0^{{\rm M}}\rangle \langle0^{{\rm M}}| \bigr]
  \nonumber\\ &\equiv&
 \sum_{p=c,d}\sum_{\ell=0,1}\prod_{\omega>0} \prod_{\bm k_\perp} \prod_{n} 
 ~{}^p\langle {\ell}_{\omega, \bm k_\perp}^{{\rm II},n}
|0^{{\rm M}}\rangle \langle 0^{{\rm M}}|{\ell}_{\omega,\bm k_\perp}^{{\rm II},n}\rangle^p,
\end{eqnarray}
which reduces to
\begin{eqnarray}
  \hat \rho^{\rm I}&=&
 \prod_{\omega>0}\prod_{~~\bm {\bm k}_\perp}\prod_{ n}\hat{\rho}^{{\rm I}}_{\omega,{\bm k}_\perp, n}
\end{eqnarray}
with 
\begin{eqnarray}
\hat{\rho}^{{\rm I}}_{\omega,{\bm k}_\perp,n}&=&
{1\over e^{-{2\pi\omega/a}}+1}
~\left(|0^{{\rm I}, n}_{\omega,{\bm {\bm k}_\perp}}\rangle^c{}^c\langle0^{{\rm I}, n}_{\omega,{\bm {\bm k}_\perp}}|+e^{-{2\pi\omega/a}}|1^{{\rm I}, n}_{\omega,{\bm {\bm k}_\perp}}\rangle^c{}^c\langle1^{{\rm I}, n}_{\omega,{\bm {\bm k}_\perp}}|\right)
\nonumber\\&&
~~~~~~~~~~~~~~\otimes
\left(|0^{{\rm I}, n}_{\omega,{\bm {\bm k}_\perp}}\rangle^d{}^d\langle0^{{\rm I}, n}_{\omega,{\bm {\bm k}_\perp}}|+e^{-{2\pi\omega/a}}|1^{{\rm I}, n}_{\omega,{\bm {\bm k}_\perp}}\rangle^d{}^d\langle1^{{\rm I}, n}_{\omega,{\bm {\bm k}_\perp}}|  \right) .
\end{eqnarray}
The Fermi-Dirac distribution function is derived as an expectation value of the number operator corresponding to particles as follows
\begin{eqnarray}
  {\rm Tr_d} [\hat{c}^{{\rm I}, n\dagger}_{\omega,{\bm k}_\perp}\hat{c}^{{\rm I}, n}_{\omega,{\bm k}_\perp} \hat \rho^{\rm I}
  ]
={1\over e^{2\pi\omega/a}+1}.
\label{Distribution I,II}
\end{eqnarray}
This is the result of the Unruh effect. 
Thus, we have explicitly demonstrated the relations for the vacuum state of 
a quantized Dirac field in 4-dimensional Minkowski spacetime covered with 
Rindler and Kasner coordinates.

\section{Summary and Conclusion}
\label{conclusion}

We investigated the solutions of a Dirac field in the four-dimensional Minkowski 
spacetime covered with the Rindler and Kasner coordinates. 
We demonstrated the construction of the mode functions in the F region, the R region, 
the L region, and the P region, and the properties of the analytic continuation of 
the positive frequency solutions are explicitly demonstrated. 
The Bogoliubov transformation between the different two sets of the mode functions is also demonstrated in the Kasner region in an explicit manner. 
The Bogoliubov transformation in the Kasner region is extended to the entire region of Minkowski spacetime, using the analytic continuation of the positive frequency solutions.
The relation between the quantum states constructed in association with the mode functions is developed, which led to the entanglement-based description of the Minkowski vacuum state. 
This description is useful to formulate the Unruh effect of the Dirac field. 
This is an extension of the previous work for a scalar field in Minkowski spacetime \cite{HIUY} 
to the case of the Dirac field. 
A unified analysis of the four-dimensional Dirac field
including the correct analytic continuation is presented for the first time.

\section*{Acknowledgement}
We thank A. Matsumura, Y. Sugiyama, H. Suzuki for valuable discussions, and we also thank Y. Kojima, Y. Nambu, J. Soda, S. Kanno, S. Iso, T. Tanaka, and S-Y. Lin for helpful comments. 
This work was supported by Japan Society for the Promotion of Science (JSPS) Grant No. 20J22946 (KU), No.~20J13640 (YN),
Japanese Government (Monbukagakusho: MEXT) Scholarship (AR), Japan Society for the Promotion of Science (JSPS) KAKENHI Grant No. 15H05895, No.~17K05444 (KY).

\appendix

\section{Matrices to relate spinor components in curved and Minkowski coordinates}
\label{Lorentz transformation of spinors in Rindler spacetime}
\begin{figure}[t]
  \begin{center}
   \includegraphics[width=120mm]{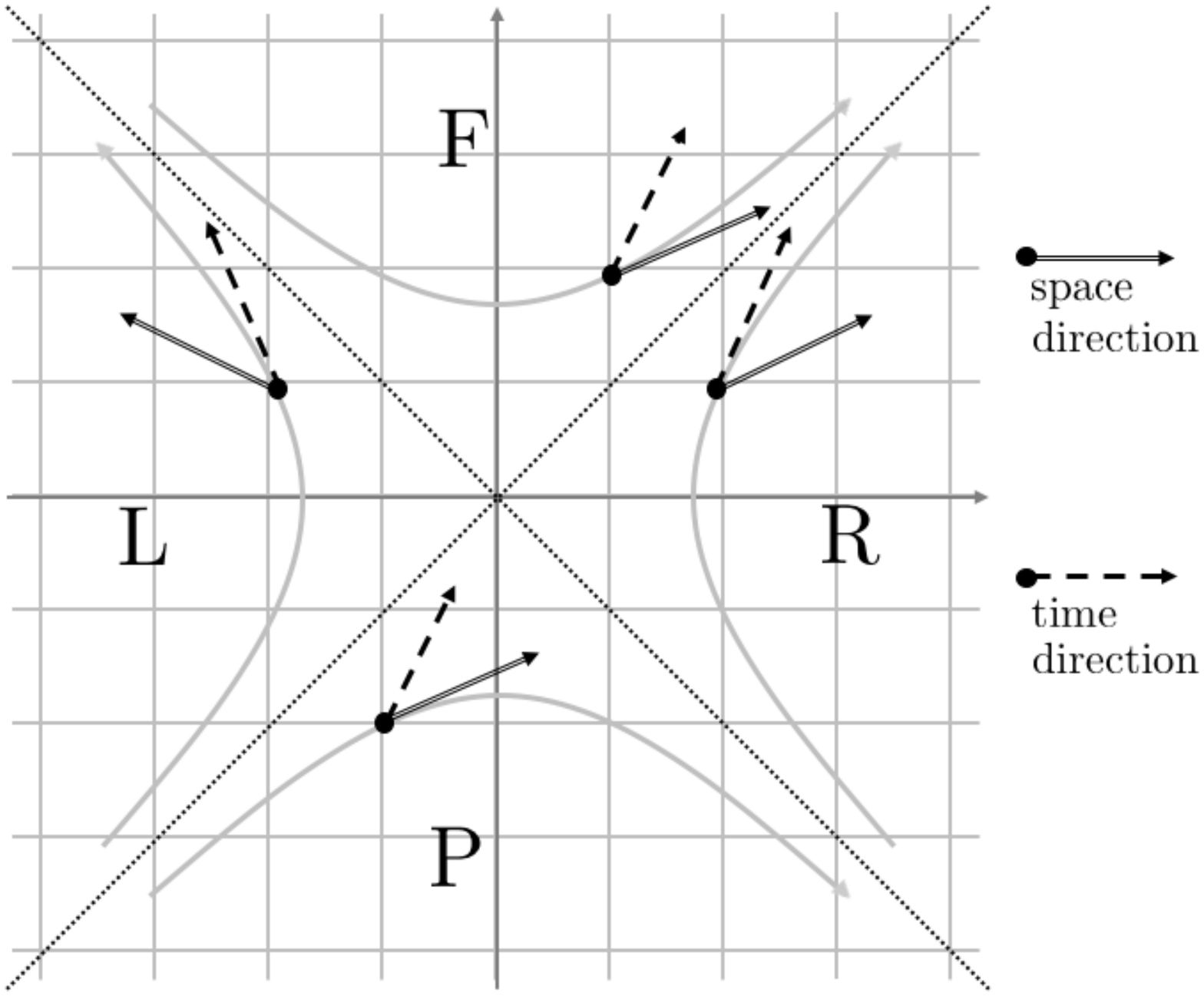}
  \caption{Local Lorentz frame in each region which indicates the deviation of direction.}
  \label{direction}
    \end{center}
\end{figure}
The spinor components are tied to the local Lorentz frame or the tetrad used to define them. To relate the
spinor solutions in the four regions found in Sec. II to one another, 
it is necessary to express them in a
common local Lorentz frame. We choose it to be the standard one, with the basis vectors pointing along with the coordinate
directions in cartesian coordinates. The tetrad in a spacetime point in each region is related to this standard tetrad
by a boost in the z-direction. Therefore, it is useful to recall how spinor components transform under such a local
Lorentz transformation. If the local Lorentz frame of the spinor components $\psi_{\rm curved}$ is obtained from those in the
standard frame, $\psi_{\rm M}$, by the boost with velocity $\tanh b$ in the $z$-direction, then
\begin{equation}
\psi_{\rm M} = \exp(b\Xi)\psi_{\rm curved} ~~{\rm or}~~\psi_{\rm curved} = \exp(-b\Xi)\psi_{\rm M},
\end{equation}
where 
\begin{equation}
\Xi = \frac{1}{4}\left[\gamma_{\rm M}^0,\gamma_{\rm M}^3\right] = \frac{1}{2}\alpha_3.
\end{equation}
The velocity in the four regions can be found in the four regions can be found from Eqs.~(\ref{Rregion}),
(\ref{Lregion}), (\ref{Fregion}) and (\ref{Pregion}) as $dz/dt$ on the timelike world line parametrized by the
time variable in each region.  They are 
$(\partial z/\partial \tau_{\rm R})/(\partial t/\partial \tau_{\rm R}) = \tanh a \tau_{\rm R}$ (R region),
$(\partial z/\partial \tau_{\rm L})/(\partial t/\partial \tau_{\rm L}) = - \tanh a \tau_{\rm L}$ (L region),
$(\partial z/\partial \eta_{\rm F})/(\partial t/\partial \eta_{\rm F}) = \tanh a\zeta_{\rm F}$ (F region)
and $(\partial z/\partial \eta_{\rm P})/(\partial t/\partial \eta_{\rm P}) = - \tanh a \zeta_{\rm P}$ (P region).
Thus, by writing the spinor solutions we found in the section II as $\psi_{\rm R}$, 
$\psi_{\rm L}$, $\psi_{\rm F}$ and
$\psi_{\rm P}$, and the corresponding ones with the standard tetrad as $\psi_{\rm M(R)}$,
$\psi_{\rm M(L)}$, $\psi_{\rm M(F)}$ and $\psi_{\rm M(P)}$, respectively, we find
\begin{eqnarray}
\psi_{\rm R}&=&\exp{(-{{a\tau_{\rm R}}\over2} \alpha_3)}~\psi_{\rm M(R)}
, \\
\psi_{\rm L}&=&\gamma^3\gamma_5\exp{({{a\tau_{\rm L}}\over2} \alpha_3)}~\psi_{\rm M(L)} 
, \\
\psi_{\rm F}&=&\exp{(-{{a\zeta_{\rm F}}\over2} \alpha_3)}~\psi_{\rm M(F)}
, \\
\psi_{\rm P}&=&\exp{({a\zeta_{\rm P}\over2}\alpha_3)}~\psi_{\rm M(P)}
.
\end{eqnarray}
For the spinors in the L region, we further need to include the matrix $\gamma^3\gamma_5$ to take into account the fact that the direction of the coordinate $\xi_{\rm L}$ in the L region is 
opposite to the one of the standard Minkowski coordinate. 

\section{Ansatz of The Minkowski Vacuum State for Fermion and Anti-fermion in Kasner Region}
\label{The ansatz of the Minkowski vacuum state for fermion and anti-fermion}
In this Appendix, we verify the validity of the ansatz for the Minkowski vacuum state 
Eq.~(\ref{Vacuum State0}) with (\ref{Vacuum State}). 
In general, we may assume the form
\begin{eqnarray}
|0^{{\rm M}, n}_{\omega,{\bm k}_\perp}\rangle =\sum_{\ell,m=0}^{1}C_{\ell m}|\ell^{{\rm K}, n(\Theta)}_{\omega,{\bm k}_\perp}\rangle^c|m^{{\rm K},\overline{n}(\Theta)}_{-\omega,-{\bm k}_\perp}\rangle^d.
\label{startp}
\end{eqnarray}
Then, the Bogoliubov transformation for the operator is expressed in the form
\begin{eqnarray}
\hat{a}^{{\rm M},n(\Theta)}_{\omega,{\bm k}_\perp}=\alpha \hat{c}^{{\rm K}, n(\Theta)}_{\omega,{\bm k}_\perp}+\beta \hat{d}^{{\rm K},\overline{n}(\Theta)\dagger}_{-\omega,-{\bm k}_\perp},
\end{eqnarray}
where $\alpha$ and $\beta$ are the non-zero Bogoliubov coefficients depending on $\omega$, ${\bm k}_\perp$ and $ n$. When we substitute the Bogoliubov transformation into the following relation,
definition of the Minkowski vacuum,  
\begin{eqnarray}
\hat{a}^{{\rm M},n(\Theta)}_{\omega,{\bm k}_\perp}|0^{{\rm M}, n}_{\omega,{\bm k}_\perp}\rangle=0,~
\label{basic}
\end{eqnarray}
with the above assumption (\ref{startp}), we get the following form
\begin{eqnarray}
\hat{a}^{{\rm M}, n (\Theta)}_{\omega,{\bm k}_\perp}|0^{{\rm M}, n}_{\omega,{\bm k}_\perp}\rangle
&=&\alpha C_{10}|0^{{\rm K},n(\Theta)}_{\omega,{\bm k}_\perp}\rangle^c|0^{{\rm K},\overline{n}(\Theta)}_{-\omega,-{\bm k}_\perp}\rangle^d+(\alpha A_{11}+\beta C_{00})|0^{{\rm K}, n(\Theta)}_{\omega,{\bm k}_\perp}\rangle^c|1^{{\rm K},\overline{n}(\Theta)}_{-\omega,-{\bm k}_\perp}\rangle^d
-\beta C_{10}|1^{{\rm K},n(\Theta)}_{\omega,{\bm k}_\perp}\rangle^c|1^{{\rm K},\overline{n}(\Theta)}_{-\omega,-{\bm k}_\perp}\rangle^d\nonumber\\
&=&0.\nonumber
\end{eqnarray}
Here we note that we have used the anti-commutation relation between 
$\hat{c}^{{\rm K},n(\Theta)\dagger}_{\omega,\bm{k}_\perp}$ and $\hat{d}^{\dagger{\rm K},\overline{n}(\Theta)}_{-\omega,-\bm{k}_\perp}$. 
Each term is expressed as an independent vector in the Fock space, therefore, 
each coefficient must be zero, and we obtain
\begin{eqnarray}
C_{10}=0~,~
C_{11}=-{\beta \over \alpha}C_{00}.
\label{C7}
\end{eqnarray}
From the property of the state of fermion,
\begin{eqnarray}
\hat{a}^{{\rm M},n(\Theta)\dagger}_{\omega,{\bm k}_\perp}|1^{{\rm M},n}_{\omega,{\bm k}_\perp}\rangle
&=&2\alpha^\ast \beta^\ast C_{01}|1^{{\rm K},n(\Theta)}_{\omega,{\bm k}_\perp}\rangle^c|0^{{\rm K},\overline{n}(\Theta)}_{-\omega,-{\bm k}_\perp}\rangle^d=0,
\nonumber
\end{eqnarray}
which yields
\begin{eqnarray}
C_{01}=0.
\end{eqnarray}
Here, we used the definition of the one particle excited state and Eq.~(\ref{startp})
\begin{eqnarray}
|1^{{\rm M},n}_{\omega,{\bm k}_\perp}\rangle 
&=&\hat{a}^{\dagger {\rm M}, n(\Theta)}_{\omega,{\bm k}_\perp}|0^{{\rm M},n}_{\omega,{\bm k}_\perp}\rangle\nonumber \\
&=&(\alpha^\ast C_{00}-\beta^\ast C_{11})|1^{{\rm K},n(\Theta)}_{\omega,{\bm k}_\perp}\rangle^c|0^{{\rm K},\overline{n}(\Theta)}_{-\omega,-{\bm k}_\perp}\rangle^d
+\alpha^\ast C_{01}|1^{{\rm K},n(\Theta)}_{\omega,{\bm k}_\perp}\rangle^c|1^{{\rm K},\overline{n}(\Theta)}_{-\omega,-{\bm k}_\perp}\rangle^d
+\beta^\ast C_{01}|0^{{\rm K},n(\Theta)}_{\omega,{\bm k}_\perp}\rangle^c|0^{{\rm K},\overline{n}(\Theta)}_{-\omega,-{\bm k}_\perp}\rangle^d.
\nonumber
\end{eqnarray}
We again used the anti-commutation relation between $\hat{c}^{{\rm K},n(\Theta)}_{\omega,\bm{k}_\perp}$ and $\hat{d}^{{\rm K},\overline{n}(\Theta)}_{-\omega,-\bm{k}_\perp}$ to derive the second equality.
Then, we obtain the ansatz for the Minkowski vacuum state, Eq.~(\ref{Vacuum State}).

\section{Explicit form of spinors in each region}
We here summarize the explicit form of the mode functions 
defined by Eqs. (\ref{PsiI,IIu}) and (\ref{PsiI,II}).
The left moving wave modes in the F-region are
analytically continued as 
\begin{eqnarray}
&&\psi^{\rm I,u}_{\omega,\bm{k}_\perp}
\equiv\left\{ \begin{array}{l}
\displaystyle 
\Psi^{\rm I,u(F)}_{\omega,{\bm k}_\perp} ~~~~~{\rm F}\\
\displaystyle 
\Psi^{\rm I,u(R/L)}_{\omega,{\bm k_\perp}} ~~{\rm R}\\
\displaystyle 
 \Psi^{\rm I,u(P)}_{\omega,{\bm k}_\perp}~~~~~{\rm P}
\end{array}
\right. ,
~~~~~~
\psi^{\rm I,d}_{\omega,\bm{k}_\perp}
\equiv\left\{
\begin{array}{l}
\displaystyle 
\Psi^{\rm I,d(F)}_{\omega,{\bm k}_\perp}~~~~~{\rm F}\\
\displaystyle 
\Psi^{\rm I,d(R/L)}_{\omega,{\bm k_\perp}}
~~{\rm R}\\
\displaystyle 
\Psi^{\rm I,d(P)}_{\omega,{\bm k}_\perp}~~~~~{\rm P}
\end{array}
\right.,
\end{eqnarray}
where these modes are zero in the L region. Their explicit forms are described as follows,
\begin{eqnarray}
&&\Psi_{\omega,{\bm k}_\perp}^{\rm I,u(F)}(x_{\rm F})=
{1\over4\pi}\sqrt{\kappa\over a}B_\omega
e^{-i\omega\zeta_{\rm F}}
e^{i{\bm k}_\perp\cdot{\bm x}_\perp}
\begin{pmatrix}
J_{-{i\omega/a}-{1/2}}(q_{\rm F})-i{m\over \kappa} J_{-{i\omega/a}+{1/2}}(q_{\rm F}) \\
i{k_1+ik_2\over \kappa} J_{-{i\omega/a}+{1/2}}(q_{\rm F}) \\
-J_{-{i\omega/a}-{1/2}}(q_{\rm F}) -i{m\over \kappa} J_{-{i\omega/a}+{1/2}}(q_{\rm F})\\
-i{k_1+ik_2\over\kappa} J_{-{i\omega/a}+{1/2}}(q_{\rm F})\\
\end{pmatrix},
\\
&&\Psi_{\omega,{\bm k}_\perp}^{\rm I,u(R/L)}(x_{\rm R})=
{1\over4\pi^2}\sqrt{\kappa\over a}B_\omega^{-1}
e^{-i\omega\tau_{\rm R}}
e^{i{\bm k}_\perp\cdot{\bm x}_\perp}
\begin{pmatrix}
K_{i\omega/a+1/2}(q_{\rm R}) +i{m\over \kappa} K_{i\omega/a-1/2}(q_{\rm R}) \\
-i{k_1+ik_2\over \kappa} K_{i\omega/a-1/2}(q_{\rm R}) \\
-K_{i\omega/a+1/2}(q_{\rm R}) +i{m\over \kappa} K_{i\omega/a-1/2}(q_{\rm R})\\
i{k_1+ik_2\over \kappa} K_{i\omega/a-1/2}(q_{\rm R})\\
\end{pmatrix}~,
\\
&&\Psi_{\omega,{\bm k}_\perp}^{\rm I,u(P)}(x_{\rm P})=
-{1\over4\pi}\sqrt{\kappa\over a}B_\omega
e^{-i\omega\zeta_{\rm P}}
e^{i{\bm k}_\perp\cdot{\bm x}_\perp}
\begin{pmatrix}
-iJ_{{i\omega/a}-{1/2}}(q_{\rm P})+{m\over \kappa} J_{{i\omega/a}+{1/2}}(q_{\rm P}) \\
-{k_1+ik_2\over \kappa} J_{{i\omega/a}+{1/2}}(q_{\rm P}) \\
iJ_{{i\omega/a}-{1/2}}(q_{\rm P})+{m\over \kappa} J_{{i\omega/a}+{1/2}}(q_{\rm P})\\
{k_1+ik_2\over\kappa} J_{{i\omega/a}+{1/2}}(q_{\rm P})\\
\end{pmatrix} , 
\end{eqnarray} 
and
\begin{eqnarray}
&&\Psi_{\omega,{\bm k}_\perp}^{\rm I,d(F)}(x_{\rm F})=
{1\over4\pi}\sqrt{\kappa\over a}B_\omega
e^{-i\omega\zeta_{\rm F}}
e^{i{\bm k}_\perp\cdot{\bm x}_\perp}
\begin{pmatrix}
-i{k_1-ik_2\over\kappa} J_{-{i\omega/a}+{1/2}}(q_{\rm F})\\
J_{-{i\omega/a}-{1/2}}(q_{\rm F})-i{m\over \kappa} J_{-{i\omega/a}+{1/2}}(q_{\rm F}) \\
-i{k_1-ik_2\over \kappa} J_{-{i\omega/a}+{1/2}}(q_{\rm F}) \\
J_{-{i\omega/a}-{1/2}}(q_{\rm F}) +i{m\over\kappa} J_{-{i\omega/a}+{1/2}}(q_{\rm F})\\
\end{pmatrix},
\\
&&\Psi_{\omega,{\bm k}_\perp}^{\rm I,d(R/L)}(x_{\rm R})=
{1\over4\pi^2}\sqrt{\kappa\over a}B_\omega^{-1}
e^{-i\omega\tau_{\rm R}}
e^{i{\bm k}_\perp\cdot{\bm x}_\perp}
\begin{pmatrix}
i{k_1-ik_2\over \kappa} K_{i\omega/a-1/2}(q_{\rm R}) \\
K_{i\omega/a+1/2}(q_{\rm R})+i{m\over\kappa} K_{i\omega/a-1/2}(q_{\rm R})\\
i{k_1-ik_2\over \kappa} K_{i\omega/a-1/2}(q_{\rm R})\\
K_{i\omega/a+1/2}(q_{\rm R})-i{m\over \kappa} K_{i\omega/a-1/2}(q_{\rm R})
\end{pmatrix}~,
\\
&&\Psi_{\omega,{\bm k}_\perp}^{\rm I,d(P)}(x_{\rm P})=
-{1\over4\pi}\sqrt{\kappa\over a}B_\omega
e^{-i\omega\zeta_{\rm P}}
e^{i{\bm k}_\perp\cdot{\bm x}_\perp}
\begin{pmatrix}
{k_1-ik_2\over\kappa} J_{{i\omega/a}+{1/2}}(q_{\rm P})\\
-iJ_{{i\omega/a}-{1/2}}(q_{\rm P})+{m\over \kappa} J_{{i\omega/a}+{1/2}}(q_{\rm P}) \\
{k_1-ik_2\over \kappa} J_{{i\omega/a}+{1/2}}(q_{\rm P}) \\
-iJ_{{i\omega/a}-{1/2}}(q_{\rm P}) -{m\over\kappa} J_{{i\omega/ a}+{1/2}}(q_{\rm P})\\
\end{pmatrix},
\end{eqnarray}
where $q_{\rm F}$, $q_{\rm R}$ and $q_{\rm P}$ are 
defined by Eqs. (\ref{qRqL}) and (\ref{qFqP}), 
$B_\omega$ is defined by Eq. (\ref{defB}), 
and $\kappa=\sqrt{m^2+{\bm k}_\perp^2}$.
The right moving wave modes in the F region are
analytically continued as follows
\begin{eqnarray}
&&\psi^{\rm II,u}_{\omega,\bm{k}_\perp}
\equiv\left\{
\begin{array}{l}
\displaystyle 
\Psi^{\rm II,u(F)}_{\omega,-{\bm k}_\perp}~~~~~{\rm F}\\
\Psi^{\rm II,u(R/L)}_{\omega,-{\bm k}_\perp}~~{\rm L}\\
\Psi^{\rm II,u(P)}_{\omega,-{\bm k}_\perp}~~~~~{\rm P}
\end{array}
\right. ,
~~~~~~
\psi^{\rm II,d}_{\omega,\bm{k}_\perp}
\equiv\left\{
\begin{array}{l}
\displaystyle 
\Psi^{\rm II,d(F)}_{\omega,-{\bm k}_\perp}~~~~~{\rm F}\\
\Psi^{\rm II,d(R/L)}_{\omega,-{\bm k}_\perp}~~{\rm L}\\
\Psi^{\rm II,d(P)}_{\omega,-{\bm k}_\perp}~~~~~{\rm P}
\end{array}
\right.  ,
\end{eqnarray}
where these modes are zero in the R region. The explicit expressions are given as follows,
\begin{eqnarray}
&&\Psi_{\omega,{-\bm k}_\perp}^{\rm II,u(F)}(x_{\rm F})=
{1\over4\pi}\sqrt{\kappa\over a}B_\omega
e^{i\omega\zeta_{\rm F}}
e^{-i{\bm k}_\perp\cdot{\bm x}_\perp}
\begin{pmatrix}
-iJ_{-{i\omega/a}+{1/2}}(q_{\rm F})+{m\over \kappa} J_{-{i\omega/a}-{1/2}}(q_{\rm F}) \\
{k_1+ik_2\over \kappa} J_{-{i\omega/a}-{1/2}}(q_{\rm F}) \\
i\kappa J_{-{i\omega/a}+{1/2}}(q_{\rm F}) +{m\over \kappa} J_{-{i\omega/a}-{1/2}}(q_{\rm F})\\
-{k_1+ik_2\over\kappa} J_{-{i\omega/a}-{1/2}}(q_{\rm F})\\
\end{pmatrix},
\\
&&\Psi_{\omega,-{\bm k}_\perp}^{\rm II,u(R/L)}(x_{\rm L})=
{1\over4\pi^2}\sqrt{\kappa\over a}B_\omega^{-1}
e^{-i\omega\tau_{\rm L}}
e^{-i{\bm k}_\perp\cdot{\bm x}_\perp}
\begin{pmatrix}
iK_{i\omega/a-1/2}(q_{\rm L}) +{m\over \kappa} K_{i\omega/a+1/2}(q_{\rm L}) \\
-{k_1+ik_2\over \kappa} K_{i\omega/a+1/2}(q_{\rm L}) \\
iK_{i\omega/a-1/2}(q_{\rm L}) -{m\over \kappa} K_{i\omega/a+1/2}(q_{\rm L})\\
-{k_1+ik_2\over \kappa} K_{i\omega/a+1/2}(q_{\rm L})\\
\end{pmatrix}~,
\\
&&\Psi_{\omega,-{\bm k}_\perp}^{\rm II,u(P)}(x_{\rm P})=
-{1\over4\pi}\sqrt{\kappa\over a}B_\omega
e^{i\omega\zeta_{\rm P}}
e^{-i{\bm k}_\perp\cdot{\bm x}_\perp}
\begin{pmatrix}
J_{{i\omega/a}+{1/2}}(q_{\rm P})-i{m\over \kappa} J_{{i\omega/a}-{1/2}}(q_{\rm P}) \\
-i{k_1+ik_2\over \kappa} J_{{i\omega/a}-{1/2}}(q_{\rm P}) \\
-J_{{i\omega/a}+{1/2}}(q_{\rm P})-i{m\over \kappa} J_{{i\omega/a}-{1/2}}(q_{\rm P})\\
i{k_1+ik_2\over\kappa} J_{{i\omega/a}-{1/2}}(q_{\rm P})\\
\end{pmatrix} ,
\end{eqnarray} %
and
\begin{eqnarray}
&&\Psi_{\omega,-{\bm k}_\perp}^{\rm II,d(F)}(x_{\rm F})=
{1\over4\pi}\sqrt{\kappa\over a}B_\omega
e^{i\omega\zeta_{\rm F}}
e^{-i{\bm k}_\perp\cdot{\bm x}_\perp}
\begin{pmatrix}
-{k_1-ik_2\over\kappa} J_{-{i\omega/a}-{1/2}}(q_{\rm F})\\
-iJ_{-{i\omega/a}+{1/2}}(q_{\rm F})+{m\over \kappa} J_{-{i\omega/a}-{1/2}}(q_{\rm F}) \\
-{k_1-ik_2\over \kappa} J_{-{i\omega/a}-{1/2}}(q_{\rm F}) \\
-iJ_{-{i\omega/a}+{1/2}}(q_{\rm F}) -{m\over\kappa} J_{-{i\omega/a}-{1/2}}(q_{\rm F})\\
\end{pmatrix},
\\
&&\Psi_{\omega,-{\bm k}_\perp}^{\rm II,d(R/L)}(x_{\rm L})=
{1\over4\pi^2}\sqrt{\kappa\over a}B_\omega^{-1}
e^{-i\omega\tau_{\rm L}}
e^{-i{\bm k}_\perp\cdot{\bm x}_\perp}
\begin{pmatrix}
-{k_1-ik_2\over \kappa} K_{i\omega/a+1/2}(q_{\rm L}) \\
-iK_{i\omega/a-1/2}(q_{\rm L})-{m\over \kappa} K_{i\omega/a+1/2}(q_{\rm L})\\
{k_1-ik_2\over \kappa} K_{i\omega/a+1/2}(q_{\rm L})\\
iK_{i\omega/a-1/2}(q_{\rm L})-{m\over \kappa} K_{i\omega/a+1/2}(q_{\rm L})
\end{pmatrix}~,
\\
&&\Psi_{\omega,-{\bm k}_\perp}^{\rm II,d(P)}(x_{\rm P})=
-{1\over4\pi}\sqrt{\kappa\over a}B_\omega
e^{i\omega\zeta_{\rm P}}
e^{-i{\bm k}_\perp\cdot{\bm x}_\perp}
\begin{pmatrix}
i{k_1-ik_2\over\kappa} J_{{i\omega/a}-{1/2}}(q_{\rm P})\\
J_{{i\omega/a}+{1/2}}(q_{\rm P})-i{m\over \kappa} J_{{i\omega/ a}-{1/2}}(q_{\rm P}) \\
i{k_1-ik_2\over \kappa} J_{{i\omega/a}-{1/2}}(q_{\rm P}) \\
J_{{i\omega/a}+{1/2}}(q_{\rm P}) +i{m\over\kappa} J_{{i\omega/ a}-{1/2}}(q_{\rm P})\\
\end{pmatrix}, 
\end{eqnarray}
where $q_{\rm L}$ is defined by Eq. (\ref{qRqL}).

\end{document}